\documentclass[aps, prl, reprint,graphicx,groupedaddress,amsmath,amssymb]{revtex4-1}
\usepackage{dcolumn}
\usepackage{bm}
\usepackage{natbib}
\usepackage{bibentry}
\usepackage[usenames,dvipsnames] {color}
\usepackage[table]{xcolor}
\usepackage{graphicx}
\usepackage{epstopdf}
\usepackage{amssymb}
\usepackage{amsmath}
\usepackage{dcolumn}
\usepackage{subfigure}
\usepackage[utf8]{inputenc}
\usepackage{amssymb}
\usepackage{amsmath}
\usepackage{chemformula}
\usepackage{chemscheme}
\usepackage[T1]{fontenc}
\usepackage{layouts}
\usepackage{siunitx}
\newcommand{\angstrom}{\textup{\AA}}

\DeclareGraphicsExtensions{.eps,.png,.pdf}

\usepackage{hyperref}
\hypersetup{
colorlinks   = true, 
urlcolor     = blue, 
linkcolor    = blue, 
citecolor   = blue 
}
\usepackage{setspace}
\usepackage[font=singlespacing,font=footnotesize, justification=RaggedRight]{caption}
\usepackage{enumitem}
\usepackage{fancybox}
\usepackage{comment}
\usepackage{url}
\usepackage{here}
\setlist{nolistsep}
\usepackage{multimedia}
\usepackage{soul}
\setstcolor{red}
\DeclareMathAlphabet      {\mathbf}{OT1}{cmr}{bx}{n}

\newcommand{\printfnsymbol}[1]{%
\textsuperscript{\@fnsymbol{#1}}%
}
\makeatother

\graphicspath{{./Figures/}{./Figures/}}

\begin{document}

\title{Scattering Evidence of Positional Charge Correlations in Polyelectrolyte Complexes}

\author{Yan Fang$^{a,b}$}
\thanks{equal contribution}
\author{Artem M. Rumyantsev$^{a,c}$}
\thanks{equal contribution}
\author{Angelika E. Neitzel$^{a,b,d}$}
\author{Heyi Liang$^{a}$}
\author{William T. Heller$^{e}$}
\author{Paul F. Nealey$^{a,b}$}
\author{Matthew V. Tirrell$^{a,b}$}
\email{mtirrell@uchicago.edu}
\author{Juan J. de Pablo$^{a,b}$}
\email{depablo@uchicago.edu}

\affiliation{
 $^a$Pritzker School of Molecular Engineering, University of Chicago, Chicago, Illinois 60637, United States \\
 $^b$Center for Molecular Engineering and Materials Science Division, Argonne National Laboratory, Lemont, Illinois 60439, United States \\
 $^c$Department of Chemical and Biomolecular Engineering, North Carolina State University, Raleigh, North Carolina 27695, United States \\
 $^d$Department of Materials Science \& Engineering, University of Florida, Gainesville, FL 32611 \\
 $^e$Neutron Scattering Division, Oak Ridge National Laboratory, Oak Ridge, Tennessee 37831, United States
}

\date{\today}

\begin{abstract}
\begin{singlespace}
Polyelectrolyte complexation plays an important role in materials science and biology. The internal structure of the resultant polyelectrolyte complex (PEC) phase dictates properties such as physical state, response to external stimuli, and dynamics. Small-angle scattering experiments with X-rays and neutrons have revealed structural similarities between PECs and semidilute solutions of neutral polymers, where the total scattering function exhibits an Ornstein-Zernike form. In spite of consensus among different theoretical predictions, the existence of positional correlations between polyanion and polycation charges has not been confirmed experimentally. Here, we present small-angle neutron scattering profiles where the polycation scattering length density is matched to that of the solvent to extract positional correlations among anionic monomers. The polyanion scattering functions exhibit a peak at the inverse polymer screening radius of Coulomb interactions, $\mathbf{q^{*} \approx 0.2~\angstrom^{-1}}$. This peak, attributed to Coulomb repulsions between the fragments of polyanions and their attractions to polycations, is even more pronounced in the calculated charge scattering function that quantifies positional correlations of all polymer charges within the PEC. Screening of electrostatic interactions by adding salt leads to the gradual disappearance of this correlation peak, and the scattering functions regain an Ornstein-Zernike form. Experimental scattering results are consistent with those calculated from the random phase approximation, a scaling analysis, and molecular simulations. 
\end{singlespace}
\end{abstract}

\maketitle

Polyelectrolyte complexes (PECs) are polymer-rich phases that result from an associative phase separation upon mixing solutions of polyanions and polycations. Polymer complexation due to electrostatic attractions is considered to be one of the physical mechanisms behind the formation of biological condensates and membrane-less organelles~\cite{brangwynne-2015, obermeyer-2020, spruijt-2021}. Within some models of prebiotic evolution, polymer-dense droplets---commonly referred to as coacervates---are proposed to have provided a means for the selective uptake of chemical species and early enzymatic activity~\cite{oparin-1938, mann-2014, keating-2019}. Synthetic PECs are promising materials with wide-ranging applications, including underwater adhesion~\cite{stewart-2011, israelachvili-2016}, gene delivery~\cite{tirrell-2021}, and separations based on nanofiltration membranes~\cite{Vos-nanofiltration}. Deciphering the role of polyelectrolyte complexation in biological processes, as well as design of PEC-based materials, requires development of a comprehensive understanding of the internal PEC structure, which governs physical properties such as surface tension and viscosity.

\begin{figure}
\centering
\includegraphics[width=0.8\linewidth]{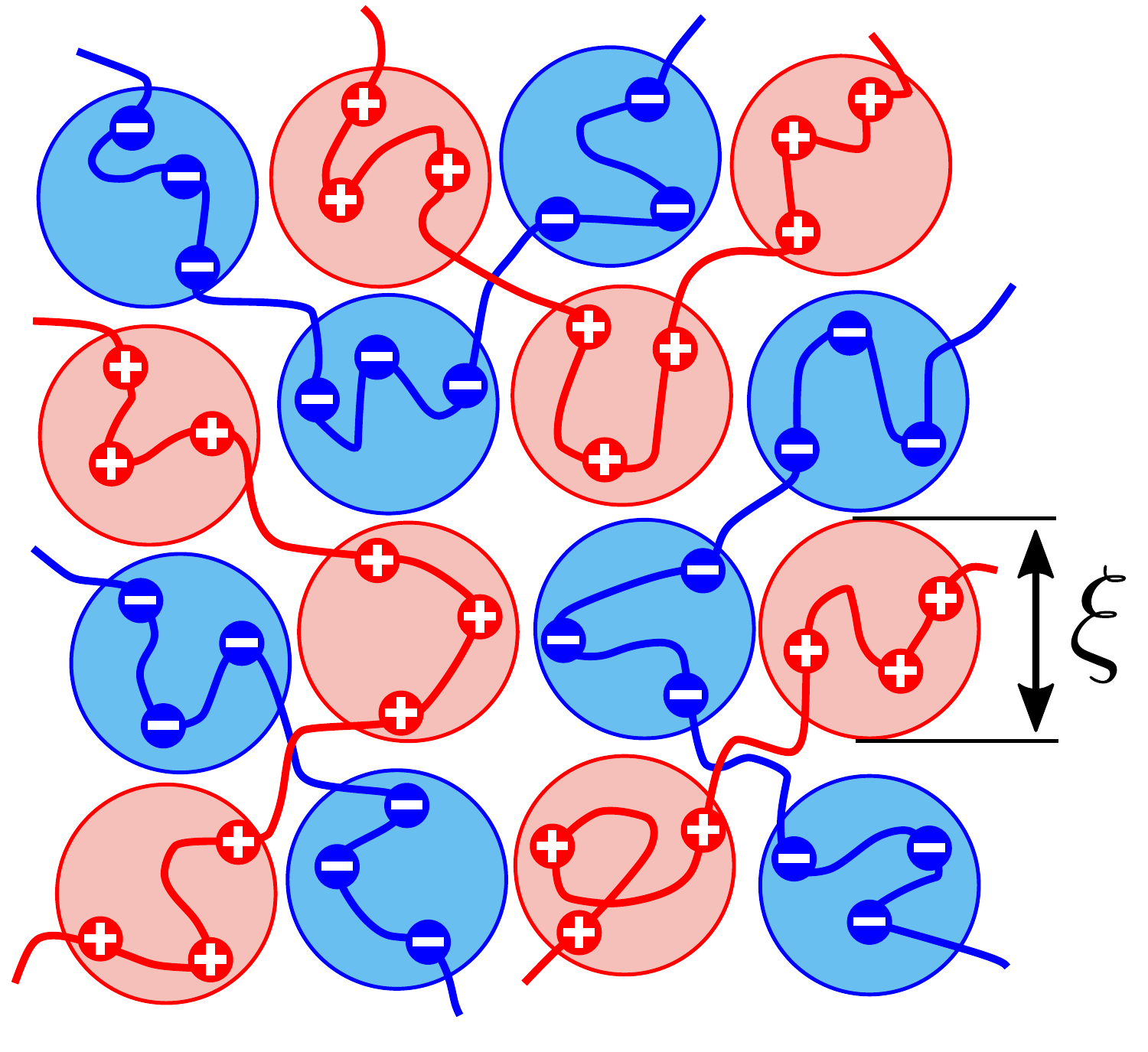}
\caption{Schematic representation of the internal structure of symmetric liquid PECs in a salt-free solvent. Polyanion ``blobs'' are mostly surrounded by polycation ``blobs'' due to the existence of charge correlations. These correlations are short-ranged and decay at a distance on the order of the correlation length, $\xi$.  }
\label{fig:blobs}
\end{figure}

A key concept pertaining to the internal structure of symmetric PECs---composed of polyanions and polycations of identical chain lengths and charge densities---is schematically illustrated in Figure~\ref{fig:blobs}. It has been suggested theoretically that~\cite{SZDR-2005, RZB-2018, rubinstein-2018} 
\begin{enumerate}
\item
Liquid PECs (i.e., polyelectrolyte complex coacervates) are structurally similar to semidilute solutions of neutral polymers;
\item
There are {\it positional correlations} in the spatial arrangement of cationic and anionic ``blobs'' (correlation volumes). Each blob is preferentially surrounded by its oppositely charged counterpart.
\end{enumerate}
In equilibrium, salt-free PECs under $\Theta$ and good solvent conditions, the size of the concentration blob (mesh size) $\xi$ is approximately equal to that of the corresponding electrostatic blob.~\cite{rubinstein-2006, RZB-2018}

Positional correlations between oppositely charged blobs are a central element of the hypothetical picture shown in Figure~\ref{fig:blobs}. As each blob is predominantly surrounded by oppositely charged neighbors, blobs experience more Coulomb attractions than Coulomb repulsions from their neighbors. For this reason, net Coulomb attractions between polyelectrolytes are referred to as correlation-induced. They provide the stability of the condensed PEC phase and are balanced by non-Coulomb short-range repulsions: two-body under athermal or good solvent conditions and three-body under $\Theta$ solvent conditions.

The arguments outlined above rely on a scaling approach. However, the first rigorous theoretical treatment of correlation-induced attractions between oppositely charged polyelectrolytes was developed via a field-theoretic approach. PECs cannot be adequately described at the level of a mean-field theory because the latter neglects correlations in the spatial arrangement of positively and negatively charged blobs. To overcome this shortcoming, Borue and Erukhimovich went beyond mean-field and applied the random phase approximation (RPA) to calculate the Gaussian fluctuation correction to the zero mean-field free energy~\cite{BE-1988, BE-1990}. Within the RPA, the effect of Coulomb interactions between polyelectrolyte charges on their density-density correlation functions is taken into account, thereby providing a reasonable description of the PEC phase. Advanced field-theoretic approaches, including field-theoretic simulations (FTS)~\cite{fredrickson-2008, fredrickson-2017} and renormalized Gaussian fluctuation theory (RGF)~\cite{wang-2018}, further support the predictions of the RPA for the structure of PECs and the charge correlations within them. We emphasize that scaling and field theory are two complementary theoretical languages that express similar physical ideas. This is demonstrated convincingly by the agreement between the analytical expressions that describe structural properties of PECs obtained from scaling arguments and the RPA~\cite{RZB-2018}. In the present work, we rely on both of these languages.

Small angle scattering experiments with X-rays (SAXS) or neutrons (SANS) are commonly employed to study the internal structure of polymer systems~\cite{degennes-book}, including that of PECs~\cite{scat-dutch-2013, scat-marciel-2018, scat-schlenoff-2018}. To date, scattering profiles of PECs have revealed structural similarities with semidilute solutions of neutral polymers; in both cases, the resultant total scattering functions exhibit an Ornstein-Zernike (OZ) form~\cite{scat-dutch-2013}
\begin{equation}
G_{tot} (q) = \frac{ G_{tot} (q=0) } { 1 + \left( q \xi \right)^{1 /\nu}},   
\end{equation}
where $1/\nu = 2$ for $\Theta$ and $1 / \nu \approx 5/3$ for good solvent conditions, respectively~\cite{scat-marciel-2018, scat-schlenoff-2018, scat-french-1975, scat-french-1976}. To the best of our knowledge, there are no reports of scattering experiments that corroborate positional charge correlations, i.e., correlations between polyanion and polycation blobs in PECs.  

In what follows, we present scattering profiles that provide direct experimental evidence in support of positional charge correlations. Specifically, SANS experiments with liquid PECs made from deuterated polycations, hydrogenated polyanions, and a solvent with a scattering length density matched to that of the polycations, reveal a peak at $q^{-1} \simeq \xi$. This result is consistent with a structure factor $G_{--} (q)$ arising from correlations between anionic blobs, as predicted by the RPA and coarse-grained molecular dynamics simulations. To further substantiate the argument that the observed peak originates from charge correlations, exogenous salt was added to the PEC phase, and the resultant SANS profiles were recorded. The RPA and simulations anticipate a substantial salt-induced screening of Coulomb interactions between electrostatic blobs, which leads to a gradual decrease of the correlation peak and the ultimate recovery of the OZ form for $G_{--} (q)$ at high salt concentrations. This evolution of the scattering profile is indeed observed in SANS experiments. Before presenting the experimental results, we review the relevant theory and discuss our simulation methods. Taken together, experiments, theory, and simulations provide an unambiguous confirmation of {\it positional charge correlations} in PECs.

\section*{Methods and Materials}

\subsection*{RPA Theory of Scattering}

We consider a symmetric stoichiometric PEC in a $\Theta$ solvent formed from long flexible polyelectrolytes, each containing a fraction $f$ of ionic monomers. The monomer size is equal to $a$, and the Bjerrum length expressed in $a$ units is denoted by $u = e^2 / \epsilon a k_B T$, where $e$ is the elementary charge and $\epsilon$ is the dielectric constant. All lengths are expressed in $a$ units and all energies in units of $k_B T$. To consider charge correlations within the salt- and counterion-free PEC, we follow a  procedure described elsewhere~\cite{BE-1988, BE-1990, joanny-2000, olvera-2004, QdP-2016, RKB-2018}, and introduce the fields of polycation and polyanion densities, $\phi_{+} (\bf{r})$ and $\phi_{-} (\bf{r})$. The free energy of the PEC includes three terms
\begin{equation}
\label{ftot-MF}
F_{tot} \left\{ \phi_{+} ({\bf r}); \phi_{-} ({\bf r})   \right\} = F_{conf} + F_{el-st} + F_{vol} 
\end{equation}
that take into account the conformational entropy of chains,  Coulomb interactions, and short-range repulsions. The first term can be written in the form of the Lifshitz entropy~\cite{GK-book}
\begin{equation}
F_{conf} = \frac{1}{6} \int \left[ \left( \nabla \sqrt{ \phi_{+} (\bf{r})} \right)^2
+ \left( \nabla \sqrt{ \phi_{-} (\bf{r})} \right)^2
\right] d^3 \bf{r}.     
\end{equation}
The electrostatic term is given by
\begin{equation}
F_{el-st} = \frac{u f^2}{2} \int 
\frac{ \left(  \phi_{+} (\bf{r}) -  \phi_{-} (\bf{r}) \right) 
\left(  \phi_{+} (\bf{r'}) -  \phi_{-} (\bf{r'}) \right) }{|\bf{r}-\bf{r'}|} 
d^3 {\bf{r}} d^3 \bf{r'}.
\end{equation}
Finally, the third contribution to the free energy is written as the leading term of a virial expansion, and corresponds to three-body repulsions  
\begin{equation}
F_{vol} = \int w \left[ \phi_{+} ( {\bf r} ) + \phi_{-} ({\bf r})  \right]^3 d^3 {\bf r}.    
\end{equation}
Here $w = C / a^6$ is the dimensionless third virial coefficient. 
By expanding the free energy into a series with respect to the density fluctuations, $\delta \phi_{\pm} ({\bf r}) = \phi_{\pm} ({ \bf r}) - \phi / 2$, and applying a Fourier transform, one obtains the following elements of the inverse matrix of the structure factors
\begin{equation}
\label{g++}
G_{++}^{-1} (q) = G_{--}^{-1} (q) = \frac{q^2}{6 \phi} + \frac{4 \pi u f^2}{q^2} + 6 w \phi     
\end{equation}
\begin{equation}
\label{g+-}
G_{+-}^{-1} (q) = G_{-+}^{-1} (q) = - \frac{4 \pi u f^2}{q^2} + 6 w \phi   
\end{equation}
Here, $\phi$ is the total polymer volume fraction within the PEC phase. The first term in eq.~\ref{g++} describes the connectivity of monomers. One can also obtain it by using the Debye function for the structure factor of ideal-coiled polyelectrolytes and neglecting the translational entropy due to their substantial length, $2 / \phi N g_{D} (x) \approx (2+x) / N \phi \approx  q^2 / 6 \phi$ where $x = N q^2 / 6$~\cite{RKB-2018, KE-1993}.   

At this point, it is convenient to introduce Edwards' correlation length
\begin{equation}
\xi_{E} = \left( 72 w \phi^2 \right)^{-1/2},
\label{xi_E}
\end{equation}
and the polymer screening radius of Coulomb interactions~\cite{BE-1988, BE-1990}
\begin{equation}
\label{r_p}
r_{p} = \left( 48 \pi u f^2 \phi \right)^{-1/4}.
\end{equation}
Inverting $G^{-1} (q)$ leads to the total structure factor of the PEC ({\it SI Appendix})
\begin{equation}
\label{gtot}
G_{tot} (q) = 2 \left( G_{++} + G_{+-} \right) 
= \frac{ \left( 6 w \phi \right)^{-1} } { 1 + \left( q \xi_{E}\right)^{2}}    \;,
\end{equation}
which corresponds to the total polymer volume fraction, $\phi ( {\bf r} ) = 
\phi_{+} ( {\bf r} ) + \phi_{-} ( {\bf r} )$, and has a simple OZ form. The respective correlation function of the total polymer density obtained via the Fourier transform, 
\begin{equation}
G_{tot} (r) = \frac{3 \phi}{ \pi r} \exp \left( - \frac{r}{\xi_{E}} \right),      
\end{equation}
has Edwards' form~\cite{degennes-book, GK-book}. The polymer charge structure factor (termed, for brevity, the charge structure factor) is given by
\begin{equation}
\label{gch}
G_{ch} (q) =  2 \left( G_{++} - G_{+-} \right) 
= \sqrt{ \frac{3 \phi} {\pi u f^2} } \;
\frac{ \left( q r_p \right)^2 } { 1 + \left( q r_p \right)^4 },
\end{equation}
and specifies the local fluctuations of the polymer charge density. It corresponds to the order parameter $ \rho ( {\bf r}) = \delta \phi_{+} ( \bf{r} ) - \delta \phi_{-} ( \bf{r} ) $, which describes segregation between polyanions and polycations~\cite{RKB-2018}. Eq.~\ref{gch} shows that $G_{ch} (q)$ has a peak at $q = r_{p}^{-1}$, which reflects the oscillatory character of the corresponding charge-charge correlation function~\cite{BE-1988, RP-2017}
\begin{equation}
\label{G-oscillating}
G_{ch} (r) = \frac{3 \phi}{\pi r} \exp \left( - \frac{r}{\sqrt{2} r_p } \right)
\cos \left( \frac{r}{\sqrt{2} r_p }  \right).
\end{equation}
This peak arises due to charge correlations between oppositely charged blobs, and is absent in $G_{tot} (q)$ given by eq.~\ref{gtot}. Fully fluctuating field-theoretic simulations (performed for diblock polyampholytes) have also predicted the existence of this correlation peak~\cite{danielsen-2019}.  

The RPA enables calculation of the Gaussian correlation correction to the free energy~\cite{BE-1990, joanny-2000}
\begin{equation}
\label{correng}
F_{corr} = \frac{1}{6 \sqrt{2} \pi r_p^3},
\end{equation}
which is responsible for attractions between the oppositely charged polyelectrolytes. This term is absent in eq.~\ref{ftot-MF}, which defines the system's free energy at the mean-field level without fluctuations. For PECs in equilibrium with the supernatant, the density is defined by the interplay between Coulomb correlation attractions and three-body repulsions
\begin{equation}
\label{eq-density}
\phi = 0.49 u^{1/3} f^{2/3} w^{-4/9}.  
\end{equation}
Strictly speaking, this result and the entire RPA analysis are valid for $w \ll 1$, when the total density fluctuations are weak. This limitation is known for neutral semidilute solutions~\cite{GK-book} and follows from a comparison of the Edwards fluctuation correction,  $F_{E} = - 1 / 12 \pi \xi_{E}^3 = - 36 \sqrt{2} w^{3/2} \phi^3 / \pi$, to the mean-field three-body term, $F_{vol} = w \phi^{3}$~\cite{fredrickson-book}. In contrast, at $w \gtrsim 1$, fluctuations are strong and scaling should be used instead of the RPA. We perform our analysis for $w \ll 1$ and provide a scaling interpretation at the crossover, $w \simeq 1$, where field-theoretic and scaling results should coincide. At $w \simeq 1$, eqs.~\ref{xi_E},~\ref{r_p}, and~\ref{eq-density} demonstrate that the polymer screening radius of Coulomb interactions and the Edwards correlation length are both on the order of the electrostatic blob size
\begin{equation}
\label{length-universality}
r_p \simeq \xi_{E} \simeq \xi_{e} \simeq u^{1/3} f^{2/3}.
\end{equation}
This equality reflects the scaling idea that, in the regime of strong fluctuations, there is a single characteristic scale called correlation length or the blob size, which quantifies both electrostatic attractions and short-range repulsions.

As salt is added to the PEC, Coulomb interactions between the polyelectrolytes are screened. The respective Debye radius decreases with increasing salt concentration, $c_s = c_{s,+} + c_{s,-}$:
\begin{equation}
\label{Debye-rad}
r_D^{-2} = 4 \pi u c_s.    
\end{equation}
At the level of the RPA, the effect of added salt can be taken into account by modifying the Fourier transform from a bare to a screened Coulomb potential, $1/q^2 \to 1/(q^2 + r_D^{-2})$, in eqs.~\ref{g++}-\ref{g+-}. In the presence of salt, the charge structure factor is equal to
\begin{equation}
\label{g++salty}
G_{ch} (q) = \sqrt{ \frac{3 \phi} {\pi u f^2} } \;
\frac{ 1 } { \dfrac{1}{Q^2 + s} + Q^2 }   
\end{equation}
with 
\begin{equation}
\label{s-param}
s = \frac{r_p^2} {r_D^2} = c_s \sqrt {\frac{\pi u } {3 \phi f^2} }; 
\qquad Q = q r_p.
\end{equation}
Following Borue and Erukhimovich~\cite{BE-1988, BE-1990}, we have introduced the reduced wavevector $Q$ and the parameter $s$ defining the reduced salt concentration. For $c_s = s =0$, eq.~\ref{g++salty} coincides with the salt-free charge structure factor given by eq.~\ref{gch}. The functional form of $G_{ch} (q)$ suggests that the scattering peak only survives at low salt concentrations, namely, at $s \leq 1$. As $c_s$ increases, the peak height goes down and the peak position shifts to lower wavevectors, $Q^{*} = \sqrt{1 - s}$. At high salt concentrations, $s \geq 1$, the peak disappears. In the limit of $s \gg 1$, when all Coulomb interactions are screened by salt, $G_{ch} (q)$ has a simple OZ form. Finally, $G_{tot} (q)$, defined by eq.~\ref{gtot}, remains unchanged upon addition of salt~\cite{BE-1988, BE-1990}.     

The results above were obtained with the standard RPA, and are sufficient to explain at a qualitative level the shape and evolution of the scattering curves obtained in experiments and simulations. Two additional modifications can be introduced to achieve quantitative agreement between theory and simulations. 

First, the effect of Coulomb correlation attractions on the correlation functions and the structure factors can be taken into account by adding the respective free energy term (Gaussian fluctuation correction) to the total free energy, $F_{tot}$~\cite{RKB-2018}. This term can be calculated within the RPA and, for salt-added systems, equals~\cite{BE-1990, joanny-2000}
\begin{equation}
F_{corr} = \frac{ (1-s) \sqrt{2-s} } {6 \sqrt{2} \pi r_p^3 }.
\label{correng-salt}
\end{equation}
Adding this term changes the Edwards screening length
\begin{equation}
\label{Edw-length}
\xi_{E}^{-2} = 12 \phi \left( 6 w \phi + \frac{d^2 F_{corr} } {d \phi^2} \right),
\end{equation}
and hence the total structure factor of the PEC
\begin{equation}
\label{gtot-final}
G_{tot} (q) = \sqrt{ \frac{3 \phi} {\pi u f^2} } \; 
\frac{1}{Q^2 + t}  \; ,
\end{equation}
while the charge structure factor remains unchanged. Here the parameter controlling the effective solvent quality is $t = r_p^2 / \xi_E^2$~\cite{BE-1988}.
With this modification, the equality between the total structure factor at $q\to 0$ and the osmotic compressibility of the PEC is now satisfied, as required by thermodynamics~\cite{degennes-book, GK-book}
\begin{multline}
\label{scat-vs-osmotic}
G_{tot} (q \to 0)
= \left[ \frac{d \left( \Pi_{vol} + \Pi_{corr} \right)} {d \phi} \right]^{-1}  \\
= \left[ \phi \frac{d^2 \left( F_{vol} + F_{corr} \right) } {d^2 \phi} \right]^{-1}.
\end{multline}
This result predicts a decrease in $G_{tot} (q = 0)$ for increasing salt concentrations when the polymer volume fraction in the PEC is constant. As salt is added, Coulomb attractions between polyelectrolytes become more screened while three-body repulsions are unchanged, leading to a decrease of PEC compressibility. Note that the approach adopted here to incorporate the effects of Coulomb correlation attractions is the simplest but not the most rigorous. For a stricter derivation of fluctuation-induced corrections to the correlation functions we refer readers to the systematic perturbation analysis by Castelnovo and Joanny~\cite{joanny-2001}.    

Second, to account for the effect of fluctuations on the charge structure factor, the Brazovskii-Fredrickson-Helfand (BFH) approximation can be used, instead of the standard RPA. The BFH approximation describes correlations in systems that may undergo order-disorder transitions~\cite{brazovskii-1975, fredrickson-1987, DE-1991, RdP-2020}. In PECs, this transition would be a microphase separation between polyanions and polycations in cases of high incompatibility. The latter can be described as short-range pairwise repulsions between polyanion and polycation monomers, $F_{rep} = \int \chi_{+-} \phi_{+} ({\bf r}) \phi_{-} ({\bf r}) d^3 r$, and $\rho (r)$ is the relevant order parameter~\cite{RKB-2018, RGK-2019, RdP-2020, fredrickson-2021, semenov-2021}. Within the mean-field calculation of the {\it correlation functions}, i.e., at the RPA level, $G_{ch} (q) \propto \left( 1/(Q^2 + s) + Q^2 - \alpha \chi_{+-} \right)^{-1} $ and the peak at $Q = Q^{*}$ in $G_{ch} (q)$ increases and diverges as the critical incompatibility is approached, $\chi_{+-}^{cr} = (2-s) / \alpha$ with $\alpha^{-1} = 2 f \sqrt{\pi u / 3 \phi}$~\cite{RKB-2018, RdP-2020}. BFH theory corrects this divergence and provides a fluctuation-induced renormalization of the charge structure factor in the disordered phase, even in the absence of polyelectrolyte incompatibility. Therefore, the charge structure factor takes the following form
\begin{equation}
\label{g++salty-BFH}
G_{ch} (q) = \sqrt{ \frac{3 \phi} {\pi u f^2} } \;
\frac{ 1 } { \dfrac{1}{Q^2 + s} + Q^2 + R}.   
\end{equation}
When expanded into a series near the peak, eq.~\ref{g++salty-BFH} reduces to the conventional BFH form~\cite{DE-1991, RdP-2020}, $G_{ch}^{-1} \propto 2-s+R + 4(1-s)\left( Q - Q^{*} \right)^{2}$. To recapitulate, the positive values of $R$ should be attributed to fluctuations, while negative $R$ are possible if the chains are effectively incompatible and $\chi_{+-} > 0$ (\textit{SI Appendix}).

\subsection*{Coarse-Grained Simulations}

Coarse-grained simulations of PECs in $\Theta$ solvent conditions were performed using an implicit solvent. Polycations and polyanions were represented by a bead-spring model ~\cite{grest-1990} each consisting of $N = 51$ beads, with a fraction $f = 1/3$ of charged beads with charge valence $z = \pm 1$. Charged beads were evenly distributed along the chain. For all PECs considered here, the polyelectrolyte concentration was kept constant and equal to that of the salt-free system. Charged beads with $z = \pm1$ were added to represent monovalent salt ions. Nonbonded interactions between beads were described by Lennard-Jones and Coulomb potentials, with parameters for the former corresponding to a $\Theta$ solvent.
Polymer beads were connected through finite-extensible nonlinear elastic (FENE) bonds. The equilibrium salt-free PEC phase was obtained in the NPT ensemble with pressure $P = 0$ representing zero osmotic pressure of polyelectrolytes in the PECs. The equilibrium polyelectrolyte concentration of the salt-free PEC was $c_{p} = c_{p,+} + c_{p,-} = 0.388 \sigma^{-3}$. Then different amounts of salt beads were added to the salt-free PEC and the new PECs were equilibrated in the NVT ensemble. The salt concentration, $c_{s} = c_{s,+} + c_{s,-}$, was varied between $0$ and $0.3 \sigma^{-3}$. All simulations were performed using LAMMPS~\cite{LAMMPS}. Details of the simulation potentials and procedures are described in the {\it SI Appendix}.

The structure factor of the PEC is defined by
\begin{equation}
G({\bf q}) = \frac{1}{ \mathcal{N}_{p}}\sum_{j=1}^{N_{p}}\sum_{k=1}^{N_{p}} \beta_{j} \beta_{k}\langle {\rm exp}[-i{\bf q}\cdot({\bf R}_{j} - {\bf R}_{k})] \rangle,
\label{structure-factor}
\end{equation}
where ${\bf q}$ is the scattering vector, $\mathcal{N}_{p}$ is the total number of polyelectrolyte beads, ${\bf R}_{j}$ is the position vector of the $j$-th bead, and $\beta_{j}$ is the form factor of the $j$-th bead. The brackets $\langle ... \rangle$ denote a statistical average. The 3D structure factor $G({\bf q})$ was evaluated for PECs equilibrated in the NVT ensemble by a combination of a fast Fourier transform (FFT) method~\cite{FFT} for $|{\bf q}| < 1.5 \sigma^{-1}$ and a direct calculation for $|{\bf q}| \geq 1.5 \sigma^{-1}$. The 1D structure factor, $G(q)$, was obtained by averaging $G({\bf q})$ with the equal magnitude of the scattering vector $q=|{\bf q}|$. The total structure factor, $G_{tot} (q)$, was calculated by setting the form factor of polycation and polyanion beads (including both charged and neutral beads) to $\beta_{+} = \beta_{-} = 1$; the polycation structure factor, $G_{++} (q)$, was calculated by setting $\beta_{+} = 1$ and $\beta_{-} = 0$; the polymer charge structure factor, $G_{ch} (q)$, was calculated by setting $\beta_{+} = 1$ and $\beta_{+} = -1$.

\subsection*{Polyelectrolyte Synthesis and Characterization}

\begin{scheme}
\centering
\includegraphics[width=0.8\linewidth]{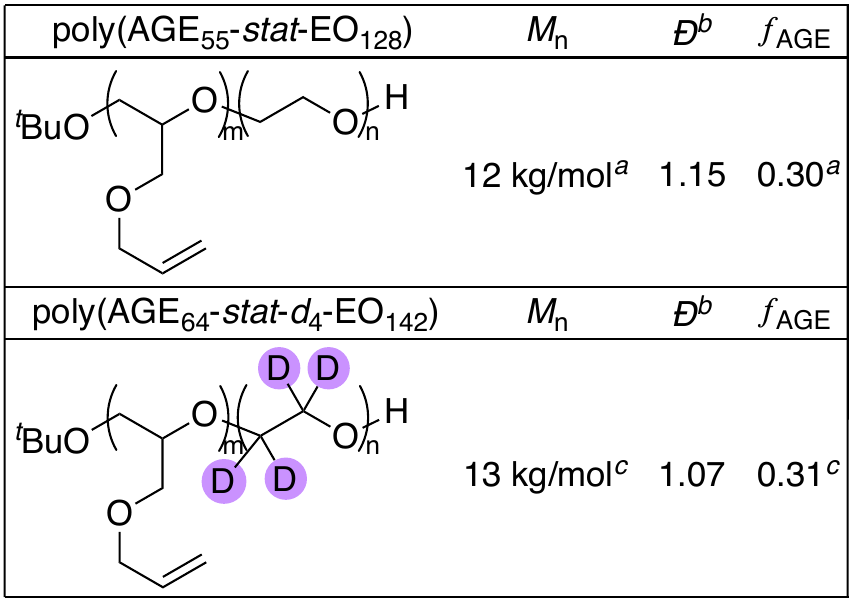}
\begin{minipage}{\columnwidth}
\caption{Molecular characteristics of near-ideally random copolyelectrolyte precursors. $^{a}$Calculated via end-group analysis using $^{1}$H NMR spectroscopy. $^{b}$ Calculated from the dRI detector signal. $^{c}$ Calculated using absolute $M_{w}$ from SEC-MALLS, {\it \DJ} from the dRI detector signal, and $N_{n, \text{AGE}}$ from $^{1}$H NMR spectroscopy with end-group analysis.}
\label{scheme:1}
\end{minipage}
\end{scheme}

Poly(allyl glycidyl ether-\emph{stat}-ethylene oxide) [poly(AGE-\emph{stat}-EO)] and poly(allyl glycidyl ether-\emph{stat}-$d_4$-ethylene oxide) [poly(AGE-\emph{stat}-$d_4$-EO)] were synthesized via oxyanionic copolymerization of allyl glycidyl ether (AGE) with ethylene oxide (EO) or ethylene-$d_4$ oxide (EO-$d_4$), respectively~\cite{neitzel-2021}. Proton nuclear magnetic resonance ($^1$H NMR) spectroscopy was used to determine the number average degree of polymerization ($N_n$) and molar fraction of AGE comonomer ($f_{\text{AGE}}$) for the protonated copolymer, and the number average degree of polymerization of AGE ($N_{\text{AGE}}$) of the partially deuterated copolymer.  

The molar mass distribution of poly(AGE-\emph{stat}-EO) was obtained from size exclusion chromatography (SEC) with dimethylformamide (DMF) as the eluent and a refractive index (RI) detector. The absolute weight average molar mass ($M_w$) of poly(AGE-\emph{stat}-$d_4$-EO) was determined by SEC with a tetrahydrofuran (THF) mobile phase using a multi-angle laser light scattering (MALLS) detector. The copolymer dispersity ({\it \DJ}) was calculated using data obtained from the RI rather than the MALLS detector. The oxyanionic polymerization of AGE suffers from chain transfer to monomer~\cite{moeller}, resulting in slight tailing towards lower molar mass chains. In this case, the {\it \DJ} obtained from MALLS is artificially narrow as the light scattering intensity is proportional to the product of concentration and $M_w$, resulting in a vanishing signal for low molar mass chains. The $N_{\text{n,total}}$ and $f_{\text{AGE}}$ for the partially deuterated copolymer could then be calculated using $M_w$ from MALLS, {\it \DJ} from the RI signal, and $N_{\text{AGE}}$ from $^1$H NMR spectroscopy (\textit{SI Appendix}, Figures S5-S7). 

Post-polymerization functionalization of neutral copolymers via reaction of allyl groups with sodium 3-mercapto-1-propanesulfonate or cysteamine hydrochloride in 4/1 dimethylformamide/water solution afforded homologous pairs of protonated or partially deuterated copolyanions and copolycations; the latter were subsequently oxidized with hydrogen peroxide~\cite{neitzel-2021}. Neutral copolymer structures and molecular characteristics are summarized in Scheme~\ref{scheme:1}. 

\begin{scheme}
\centering
\includegraphics[width=\linewidth]{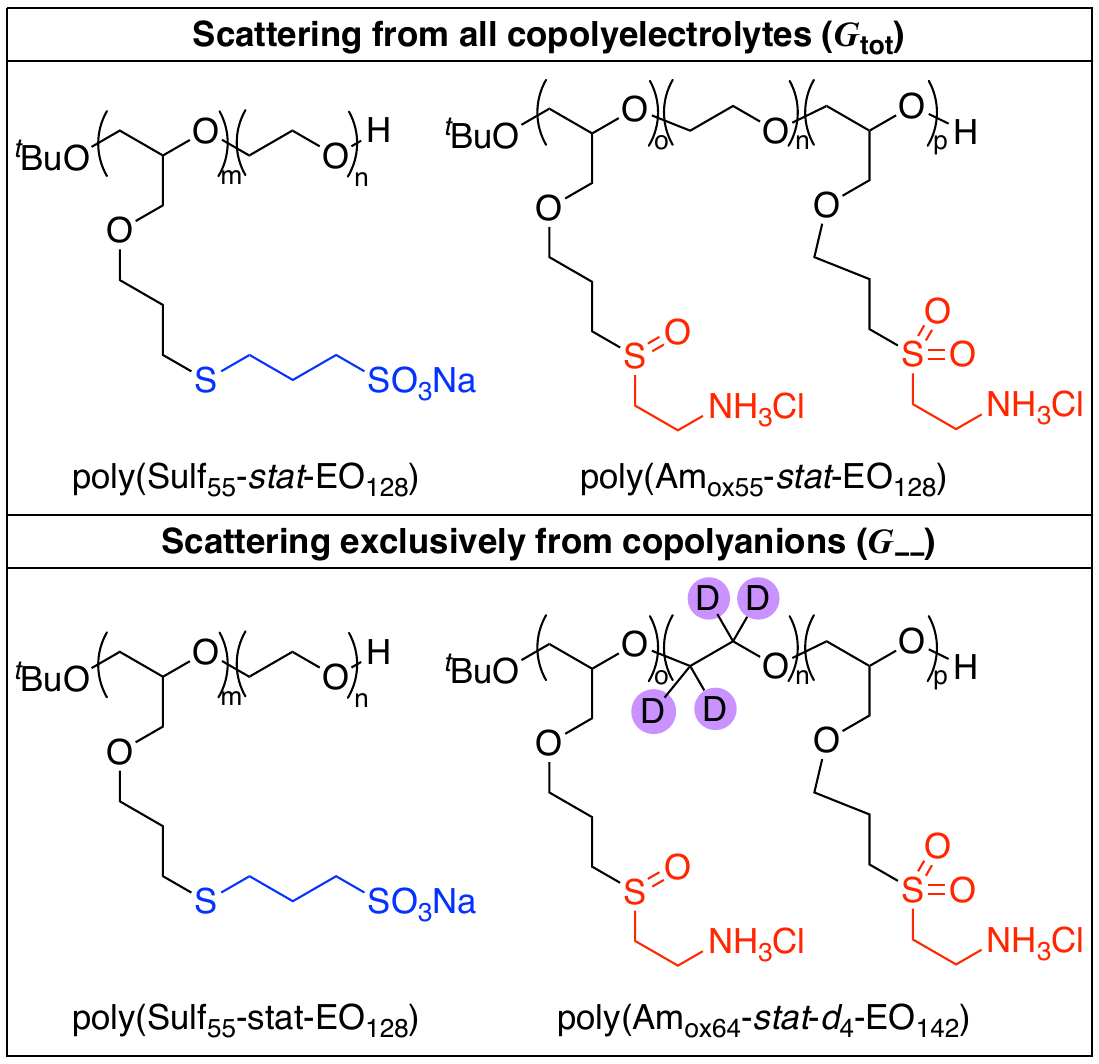}
\caption{Structures of near ideally random copolyelectrolytes and combinations thereof to measure the total polymer structure factor, $G_{tot} (q)$, and structure factor of polyanions, $G_{--} (q)$. }
\label{scheme:2}
\end{scheme}

\subsection*{Preparation of Polyelectrolyte Complexes}

Samples were prepared on a 45~mL scale, distributed into 3x15~mL centrifuge tubes and centrifuged at $5000 \times g$ for 90 min. All samples were prepared at a charge-stoichiometric ratio and a polymer concentration of $c_{P,i}$ = 5~mg/mL (0.5 wt \%). The supernatant was removed and the PEC phases from three samples were combined by weight in a tared Hellma quartz cuvette (path length = 1 mm) for measurement. The combined mass of each PEC sample was $\approx 350$~mg. For samples prepared with exogenous salt, sodium chloride (NaCl) was weighed into a tared 1.5~mL Eppendorf tube, the PEC sample prepared as described before, and transferred into the Eppendorf tube. The sample was then vortexted prior to transferring it into the cell. Contrast matching conditions were calculated and measured experimentally for each polyelectrolyte (\textit{SI Appendix}, Table S2 and Figure S11). 

\subsection*{SANS Experiments and Data Analysis}

SANS experiments were performed at the Extended Q-Range Small-Angle Neutron Scattering Diffractometer (EQ-SANS) at the Spallation Neutron Source (SNS) at Oak Ridge National Laboratory (ORNL)~\cite{heller-2018}. To collect the SANS data presented here, two configurations of the instrument were employed: 4~m sample-to-detector distance with a wavelength band of min = 2.5~$\angstrom$ and 2.5~m sample-to-detector distance with min = 2.5~$\angstrom$. This provided momentum transfer $q$, $q = 4 \pi \sin(\theta) / \lambda$, from 0.01~$\angstrom^{-1}$ to 0.70~$\angstrom^{-1}$; here $2 \theta$ is the scattering angle and $\lambda$ is the neutron wavelength. All data reduction was accomplished using the {\it drtsans} software~\cite{heller-2022}, which performs the standard data reduction corrections such as for wavelength-dependent flux and neutron transmission, detector response, empty cell scattering, etc. We used a calibrated porous silica standard to get the scale factor to convert the measured data into absolute intensity units of 1/cm~\cite{bates-1987}. The temperature was kept constant at $20.0 \pm 0.1 ^\circ$C using one of the standard sample environments of the instrument to make the measurement temperature identical to the temperature at which the samples were originally prepared.

Scattering profiles were obtained for 1) poly(Sulf-\emph{stat}-EO)/poly(Am$_{\text{ox}}$-\emph{stat}-EO) in \ch{D2O}, 2) poly(Sulf-\emph{stat}-$d_4$-EO))/poly(Am$_{\text{ox}}$-\emph{stat}-EO) matching the scattering length density (SLD) of the solvent to that of the polyanion, and 3) poly(Sulf-\emph{stat}-EO)/poly(Am$_{\text{ox}}$-\emph{stat}-$d_4$-EO) matching the SLD of the solvent to that of the polycation. These provided the experimental profiles of G$_{tot}$, G$_{++}$, and G$_{--}$. The signal for G$_{++}$ was weak due to increased incoherent scattering from water and thus further measurements were confined to those of G$_{--}$. The reduced data were plotted using the Irena SAS package in IGOR Pro and solvent backgrounds were subtracted from PEC scattering profiles. Charge structure factor profiles, $G_{ch} (q)$, were generated by overlapping the experimental $G_{--} (q)$ and $G_{tot} (q)$ profiles at high $q$, normalizing their values to provide twice higher values of $G_{tot} (q)$ in this $q$-region, and then subtracting the scattering profiles from each other according to the theoretical equation $G_{ch} (q) = 4 G_{--} (q) - G_{tot} (q)$.

\section*{Results and Discussion}

\subsection*{Choice of Experimental System}

The theoretical picture provided by the RPA and scaling analysis was developed for weakly charged PECs at equilibrium. Polyelectrolyte complexation can produce liquid- or solid-like PECs; the latter are known to be kinetically trapped structures below their glass transition temperature. To properly corroborate the theoretical picture, we chose polyelectrolytes derived from poly(allyl glycidyl ether-\textit{stat}-ethylene oxide) [poly(AGE-\textit{stat}-EO)], which yield polyelectrolyte complexes of liquid-like character at a charge density of $f = 0.3$.~\cite{neitzel-2021} Aqueous solubility at low charge density is facilitated by incorporation of EO as the neutral comonomer. AGE was chosen as the precursor to either the cationic or anionic monomer. Postpolymerization modification to polyelectrolytes from a common neutral precursor eases polymer characterization by conventional methods and produces homologous polycation/polyanion pairs, which reduces the number of variables to consider downstream. As EO is commercially available with near quantitative deuterium labeling, this platform enables the synthesis of an analogous, deuterated polycation/polyanion pair by the same synthetic methodology.

\subsection*{Salt-Free Polyelectrolyte Complexes}

\begin{figure}[h]
\centering
\includegraphics[width=0.8\linewidth]{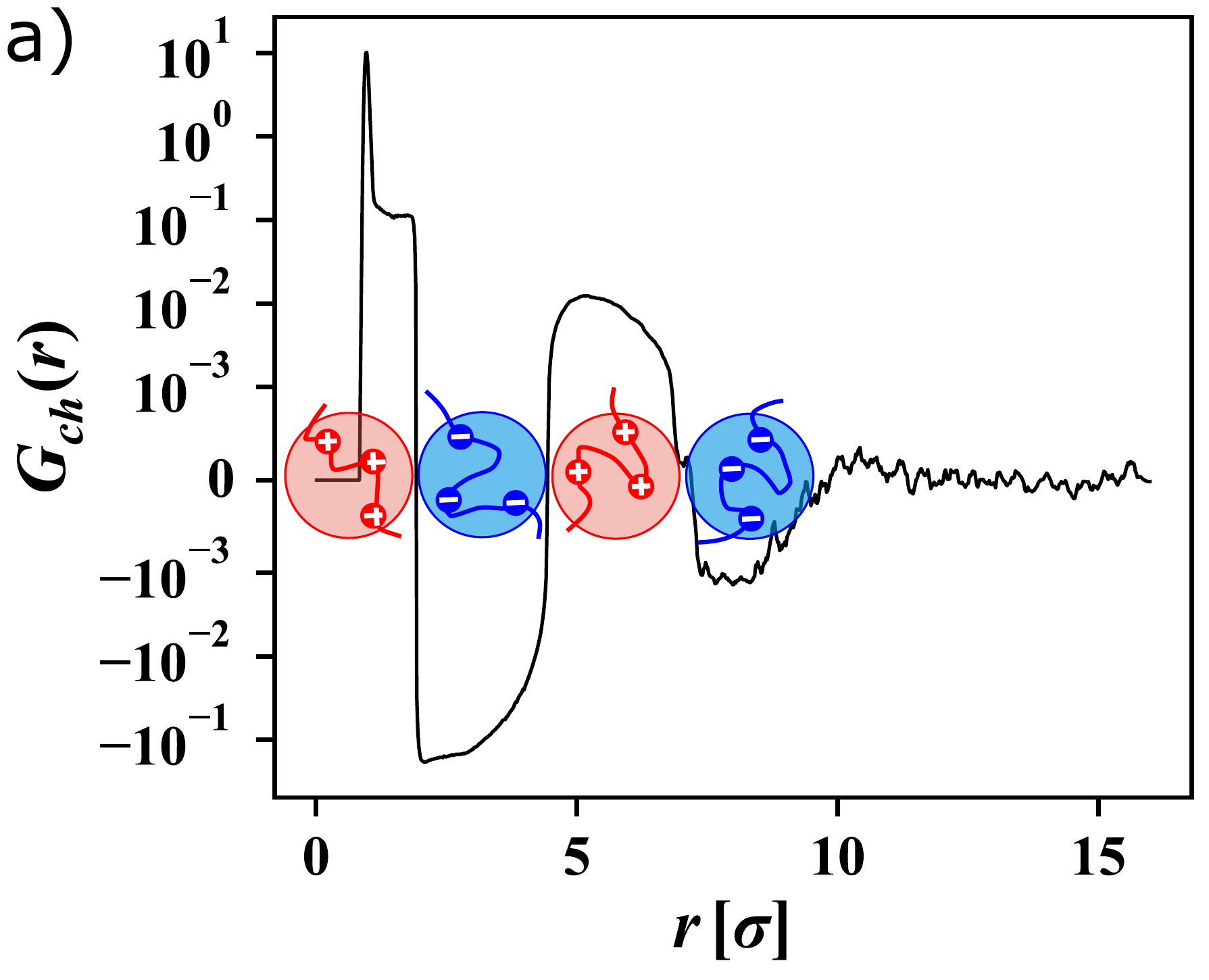}
\includegraphics[width=0.8\linewidth]{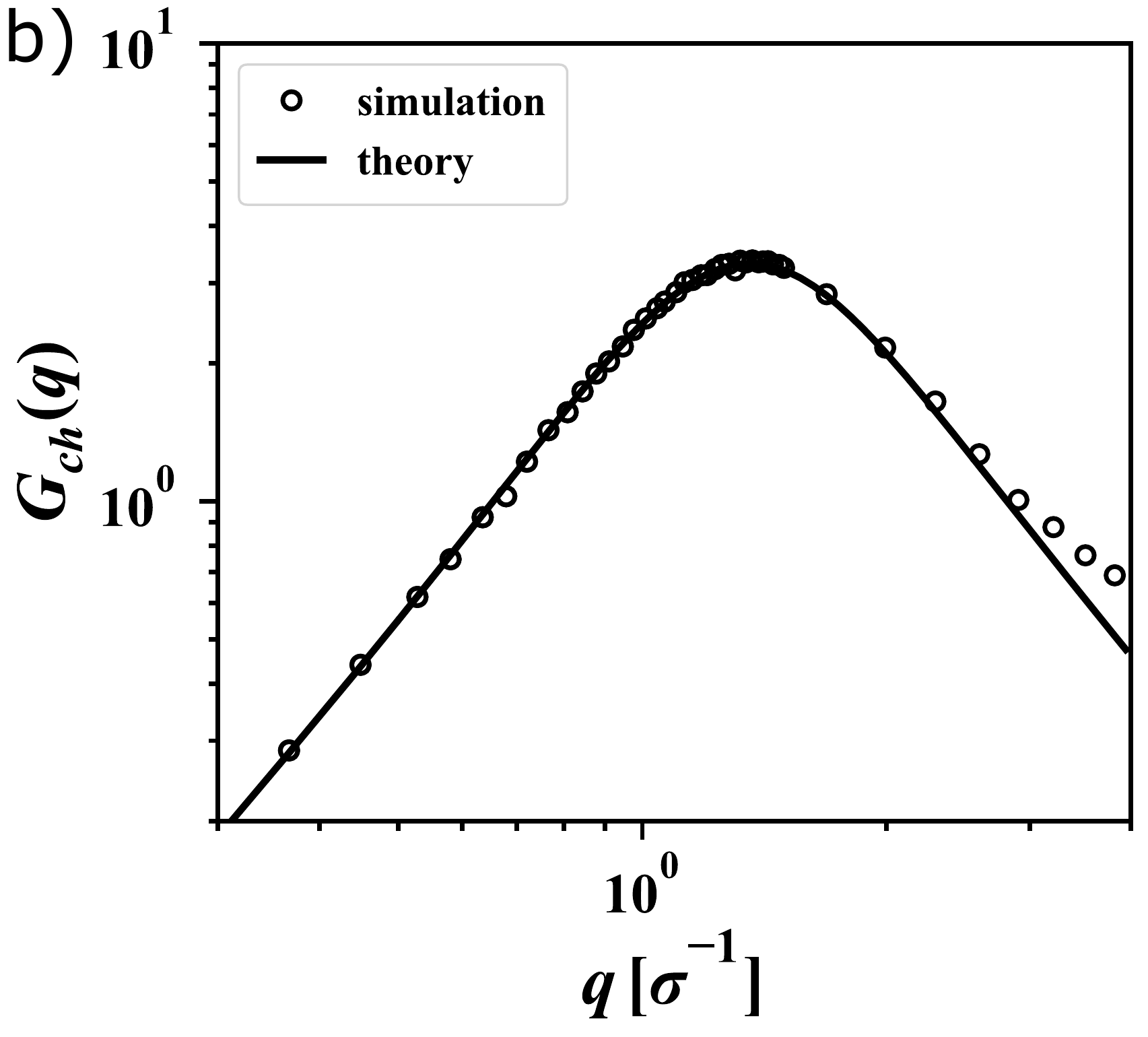}
\caption{ (a) Radial distribution function of charge, $G_{ch} (r)$, and (b) charge structure factor, $G_{ch} (q)$, of the salt-free PEC in a $\Theta$ solvent from coarse-grained simulations. $G_{ch} (q)$ corresponds to the Fourier transform of $G_{ch} (r)$, eq~\ref{FT}. In Figure a, the $y$ axis is plotted on a symmetric log scale, while in a linear scale within ($-10^{-3}$,$10^{-3}$). In Figure b, the solid line is the best theoretical fit for the simulation data (open circles) with the functional form given by eq.~\ref{gch-fit}. In the coarse-grained simulations, the chain length is $N = 51$, polyelectrolytes contain a fraction $f = 1/3$ of ionic monomers, and the Bjerrum length is set to $l_B / \sigma = 1$.}
\label{fig:RDF}
\end{figure}

We start our analysis of charge correlations in equilibrium, salt-free PECs in a $\Theta$ solvent with results of coarse-grained simulations. The radial distribution function (RDF) of charges $G_{ch} (r)$, plotted in Figure~\ref{fig:RDF}a, exhibits a damped oscillatory behavior. Owing to the zero average polymer charge density within the PEC phase, the charge RDF coincides with the charge-charge correlation function. The latter can be calculated within the RPA, eq.~\ref{G-oscillating}, and predicted to oscillate~\cite{BE-1988, RP-2017}. These oscillations are absent in the comparative system of disjointed charges, where charge screening has a simple Debye-H\"{u}ckel form. They are a unique feature of polyelectrolyte systems, which arises from the interplay between the connectivity of charges and the long-range Coulomb interaction potential~\cite{khokhlov-1982, BE-1988}. The RPA shows that the period of oscillations is equal to $D = 2\sqrt{2} \pi r_p$, where $r_p$ is the polymer screening radius of Coulomb interactions in the PEC, given by eq.~\ref{r_p}~\cite{BE-1988, RP-2017}. At equilibrium, in PECs with $w \simeq 1$, the polymer screening radius is on the order of the electrostatic blob size and the concentration correlation length, $r_p \simeq \xi_e \simeq \xi$. This enables a simple scaling interpretation of the oscillations in terms of the spatial correlations of oppositely charged blobs~\cite{remark-1}, as shown schematically in Figure~\ref{fig:RDF}a. 

\begin{figure}
\centering
\includegraphics[width=0.8\linewidth]{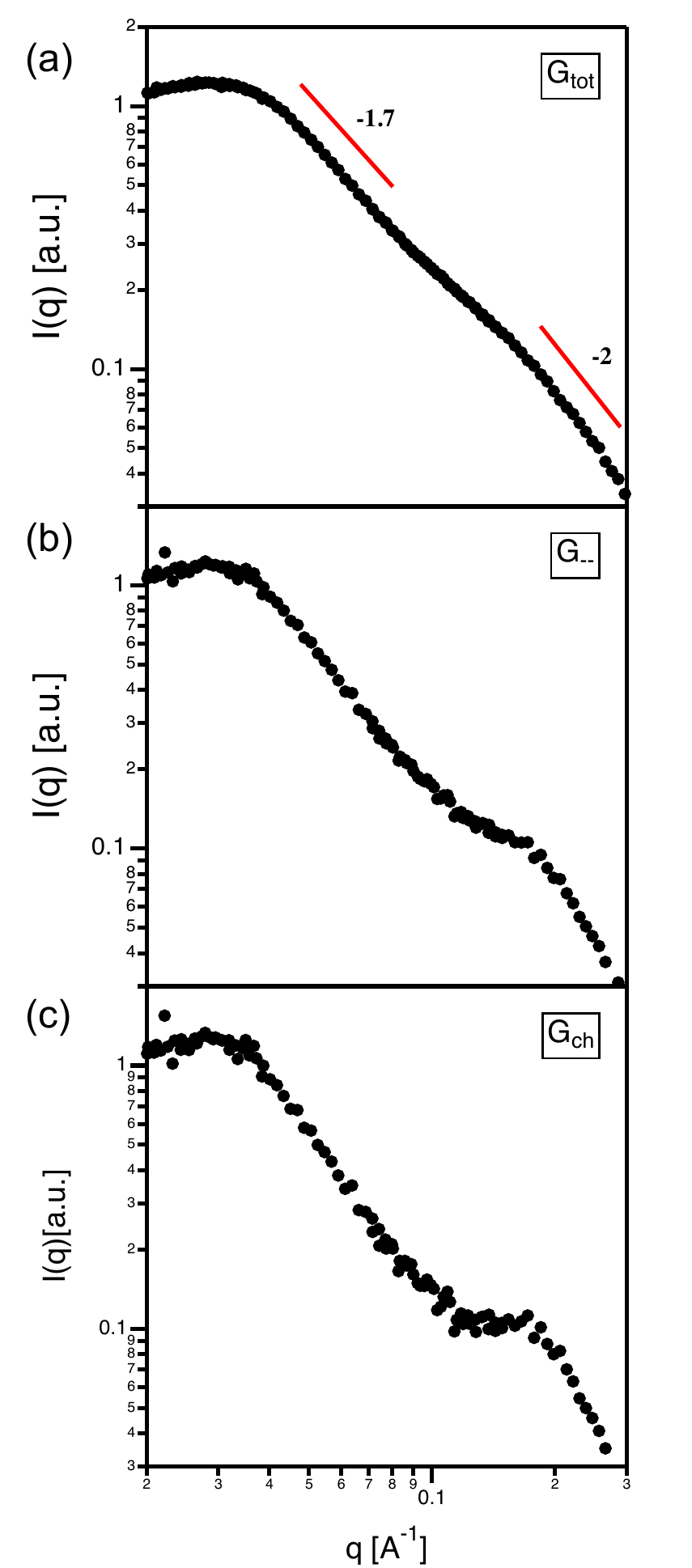}
\caption{Background-subtracted scattering intensity $I (q)$ as a function of the wavevector $q$ obtained from SANS experiments with PECs of charge fraction $f = 0.30$ without exogenous salt. 
(a) $G_{tot}$ is obtained via complexation of fully protonated chains of 
poly(Am$_{\text{ox}}$-{\it stat}-EO) and poly(Sulf-{\it stat}-EO) and measured against \ch{D2O};
(b) $G_{--}$ is obtained via complexation of fully protonated poly(Sulf-{\it stat}-EO) and partially deuterated poly(Am$_{\text{ox}}$-{\it stat}-$d_4$-EO)
and measured against \ch{H2O} / \ch{D2O} $= 49/51$ that contrast matches deuterated polycation;
(c) $G_{ch}$ is obtained by subtracting $G_{tot}$ from $G_{--}$ through the theoretical equation $G_{ch} = 4 G_{--} - G_{tot}$ after their appropriate normalization at high $q$ values.}
\label{fig:expscat}
\end{figure}

The periodicity of the function in real space leads to the peak at $q^{*} \simeq 2 \pi / D$ in the Fourier representation. For undamped harmonic functions---such as in systems with long-range order---this peak has the shape of a $\delta$-function, whereas exponential damping of oscillations is manifest in the peak's broadening and a slight shift in position. The latter is observed in the Fourier transform of the charge-charge correlation function
\begin{equation}
\label{FT}
G_{ch} (q) = \frac{1}{ \left( 2 \pi \right)^3 } 
\int G_{ch} ( {\bf r}) e^{i {\bf q r}} d^3 {\bf r},
\end{equation}
which was calculated in simulations and is shown in Figure~\ref{fig:RDF}b. The position of the correlation peak, $Q^{*} \ll q_{m} = 2 \pi / \sigma$, is indicative of length scales substantially larger than the size of the monomer, $\sigma$. The monomer peak due to the covalent connectivity of beads at much higher wavevectors, $q_{m}$, is observed in simulations (\textit{SI Appendix}, Figure S1) and falls outside of the range of the angles probed experimentally. The charge correlation peak is theoretically anticipated within the RPA~\cite{BE-1988, BE-1990}, eq.~\ref{gch}. Introducing the dimensionless wavevector $Q = q r_p$, the theoretical functional form of the charge structure factor is predicted to be
\begin{equation}
G_{ch} (Q) \propto \frac{1}{ \dfrac{1}{Q^2} + Q^2 }.   
\label{Gch-sf-simple}
\end{equation}
By minimizing the denominator one can see that the peak in $G_{ch} (q)$ is expected at $Q^{*} = 1$, i.e., at $q^{*} = r_p^{-1} \simeq 1/D \simeq \xi$, in agreement with the period of oscillations in the respective charge correlation function, eq.~\ref{G-oscillating}. To recapitulate, both the RPA and coarse-grained simulations predict the existence of the scattering peak in the charge structure factor, corresponding to
the existence of positional correlations between the monomers of polyanions and polycations.

\begin{figure*}
\centering
\includegraphics[width=17.8cm]{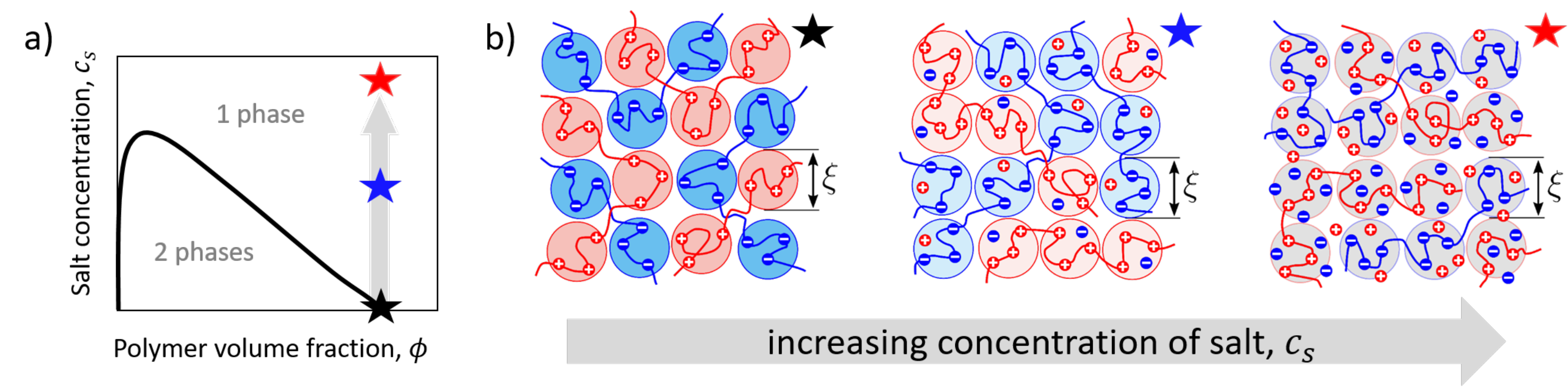}
\caption{ (a) Schematic representation of the binodal phase diagram (black line) and compositions of the PECs with fixed polymer density and varying salt content (stars). 
(b) Evolution of the blob structure of the respective PECs upon addition of salt. The size of the concentration blob $\xi$ remains unchanged owing to the constant density of the PEC. }
\label{fig:chart}
\end{figure*}

Figure~\ref{fig:expscat} shows SANS profiles for "salt-free" PECs of poly(Sulf$_{55}$-{\it stat}-EO$_{128}$)/poly(Am$_{\text{ox}55}$-{\it stat}-EO$_{128}$) and poly(Sulf$_{55}$-{\it stat}-EO$_{128}$)/poly(Am$_{\text{ox}64}$-{\it stat}-$d_4$-EO$_{142}$) (Scheme~\ref{scheme:2}) with structure factors $G_{tot} (q)$, $G_{--} (q)$, and $G_{ch} (q)$. "Salt-free" PEC refers to samples prepared without exogenous salt; hence they contain a fraction of the counterions introduced by the polyelectrolytes. Analogous PEC samples prepared at $c_{P,i}$ = 10 mg/mL were previously shown to contain $c_{s,PEC}$ = 0.17 wt \% NaCl~\cite{neitzel-2021}. SANS samples were prepared at $c_{P,i}$ = 5 mg/mL on a scale of ca. 350 mg. It is estimated that "salt-free" PEC samples in this work contain less than $0.0017 \times 350$ mg $= 0.6$ mg NaCl as it has been shown that decreasing the $c_{P,i}$ results in denser PECs (i.e., higher $c_{P,PEC}$) that partition less salt (i.e., lower $c_{s,PEC}$)~\cite{Li-2018}. The total structure factor, $G_{tot} (q)$, obtained by combining non-deuterated polyelectrolytes in \ch{D2O}, displays the expected OZ form, consistent with earlier reports~\cite{scat-dutch-2013, scat-marciel-2018, scat-schlenoff-2018} and eq.~\ref{gtot-final}, reaffirming the structural semblance of liquid PECs and semidilute solutions of neutral polymers. A gradual slope change from $-1.7$ to $-2$ in the intermediate $q$-range suggests that chain conformations in the PEC phase are close to the crossover between good and $\Theta$ solvent quality for PE chains, i.e., between Gaussian coil and self-avoiding walk statistics~\cite{RZB-2018}. The theoretical and simulation results discussed earlier were presented for the case of $\Theta$ solvent, as the main conclusions of this work do not change when considering a good/athermal solvent. The exact RPA functional form and complimentary simulation scattering profiles for $G_{ch} (q)$ in the case of an athermal solvent are provided in SI Appendix.

The form of the polyanion structure factor, $G_{--} (q)$, is more complicated. A clear shoulder emerges at $q^{*} \approx 0.2~\angstrom^{-1}$, which we attribute to positional charge correlations. A similar shoulder was reported in ref.~\citenum{scat-schlenoff-2018} in solid-like PECs. The origin of the peak can be understood by considering the relationship between the different structure factors
\begin{equation}
\label{relation-of-gs}
G_{--} (q) = \frac{ G_{tot} (q) + G_{ch} (q) }{4}.
\end{equation}
$G_{tot} (q)$ exhibits an OZ form that decreases monotonically, so one can argue that the shoulder in $G_{--} (q)$ can only appear due to the peak in $G_{ch} (q)$. The shoulder in $G_{--} (q)$ is seen in simulations (black curve in Figure~\ref{fig:theorscat}b), consistent with a previous report~\cite{dobrynin-2021}. We emphasize that eq.~\ref{relation-of-gs} is general and independent of the type of theory used to describe scattering from PECs. It follows from the definition of the structure factors and the stoichiometry of the PEC, $\phi_{+} = \phi_{-}$, which is provided by the experimental procedure to prepare the PECs. The resulting $G_{tot} (q)$ and $G_{--} (q)$ permits direct calculation of the charge structure factor $G_{ch} (q)$ (see experimental SANS section), which is shown in Figure~\ref{fig:expscat}c. The correlation peak in $G_{ch} (q)$ is more evident than in $G_{--} (q)$, as is expected from the theory. The position of the peak is unchanged, $q^{*} \approx 0.2~\angstrom^{-1}$, which corresponds to a polymer screening radius for Coulomb interactions equal to $r_p = 1/q^{*} \simeq 0.5$ nm and a period of oscillations in the charge-charge correlation function equal to $D = 2 \sqrt{2} \pi r_p \approx 4.5$ nm. The latter can be viewed as an estimate for the size of the electrostatic blob.

Fitting $G_{tot} (q)$ with the OZ functional form given by eq.~\ref{gtot} allows one to obtain another estimate of the blob size or, rigorously speaking, the Edwards correlation length. At $q_E = \xi_E^{-1} \approx 0.06~\angstrom^{-1}$, the value of $G_{tot} (q)$ is half of the plateau value, which yields $\xi_E \approx 1.7$ nm. Thus, in the experiments, the polymer screening radius of Coulomb interactions and the Edwards screening length of the excluded volume interactions differ by a factor of more than three.
This illustrates the approximate nature of the scaling relationship $r_p \simeq \xi_E  \simeq \xi$ and the qualitative nature of the scaling arguments if applied at $w \ll 1$.  
In fact, the scaling approach is unable to predict the dependence of PEC density and correlation length on $w$~\cite{RZB-2018}; for this reason, it does not distinguish between $r_p$ and $\xi_E$. Within the field theory, the difference between $r_p$ and $\xi_E$ defines the value of the parameter describing the effective solvent quality for the PEC, $t = r_p^2 / \xi_E^2$, which is equal to $t \approx 0.09$ in experiments. The fact that $t \ll 1$ is consistent with theoretical expectations and estimates from simulations. Theoretically, the equilibrium density of the salt-free PEC is defined by eq.~\ref{eq-density}, which yields a polymer screening radius equal to $r_p = 0.34 u^{-1/3} f^{-2/3} w^{1/9}$ (see eq.~\ref{r_p}). The Edwards correlation length is given by eq.~\ref{Edw-length}. Using eq.~\ref{correng} for the free energy correlation term, one can find $\xi_E = 0.28 u^{-1/3} f^{-2/3} w^{-1/18}$ and $t = 1.47 w^{1/3}$. Since for all real polymers the dimensionless third virial coefficient is low, $w \ll 1$~\cite{GK-book}, the value of $t$ should be substantially lower than unity. This conclusion is also supported by our simulations where, for salt-free PECs, we used the theoretical functional forms of eqs.~\ref{gtot-final} and~\ref{g++salty-BFH} to fit the total and charge structure factors and found $r_p = 0.74 \sigma$, $\xi_{E} = 1.42 \sigma$, and $t = 0.27$.

It can be seen that the experimental $G_{ch} (q)$ does not tend to zero at low $q$ (Figure~\ref{fig:expscat}c), contrary to what is predicted by theory and simulations. As mentioned previously, "salt-free" PECs contain a low, yet non-negligible concentration of counterions, which screen Coulomb interactions. This implies that eq.~\ref{Gch-sf-simple} is not exact for PECs prepared without exogenous salt. Eq.~\ref{g++salty} with non-zero reduced concentration of salt, $s \neq 0$, should be used instead, which releases the $G_{ch} (q = 0) \to 0$ requirement. 
Additionally, we speculate that differences between theory and experiment arise due to deviations from ideality that are inevitable in real experimental systems. Even when polyanions and polycations are derived from an identical precursor, chemical differences between charged groups translate into polyelectrolytes with unequal SLDs. Deuterated polycations and hydrogenated polyanions were necessarily derived from different precursors and therefore do not have {\it identical} chain lengths and charge fractions (Scheme ~\ref{scheme:1}). Collectively, this makes eq.~\ref{relation-of-gs} only approximately valid for calculations of the experimental $G_{ch} (q)$. Lastly, a more pronounced difference in the SLDs of hydrogenated polyanions and deuterated polycations is desirable; however, this would require full deuteration of polycations, which is synthetically challenging (see Scheme~\ref{scheme:2}).

\subsection*{Effect of Salt}

To corroborate that the peak in the experimental scattering functions arises from charge correlations we examined the effect of exogenous salt. Salt is expected to screen charge correlations and thereby lead to the gradual disappearance of the correlation peak. In order to decouple the effects of salt and polymer concentrations, the latter was fixed to that of the "salt-free" PEC, whereas the salt concentration was varied as shown with stars in Figure~\ref{fig:chart}a. Importantly, the salted PECs are not in equilibrium with a supernatant phase and do not coexist with it. In experiments, this was realized by adding salt to the equilibrated, "salt-free" PECs after removal of the supernatant.

\begin{figure}
\centering
\includegraphics[width=0.8\linewidth]{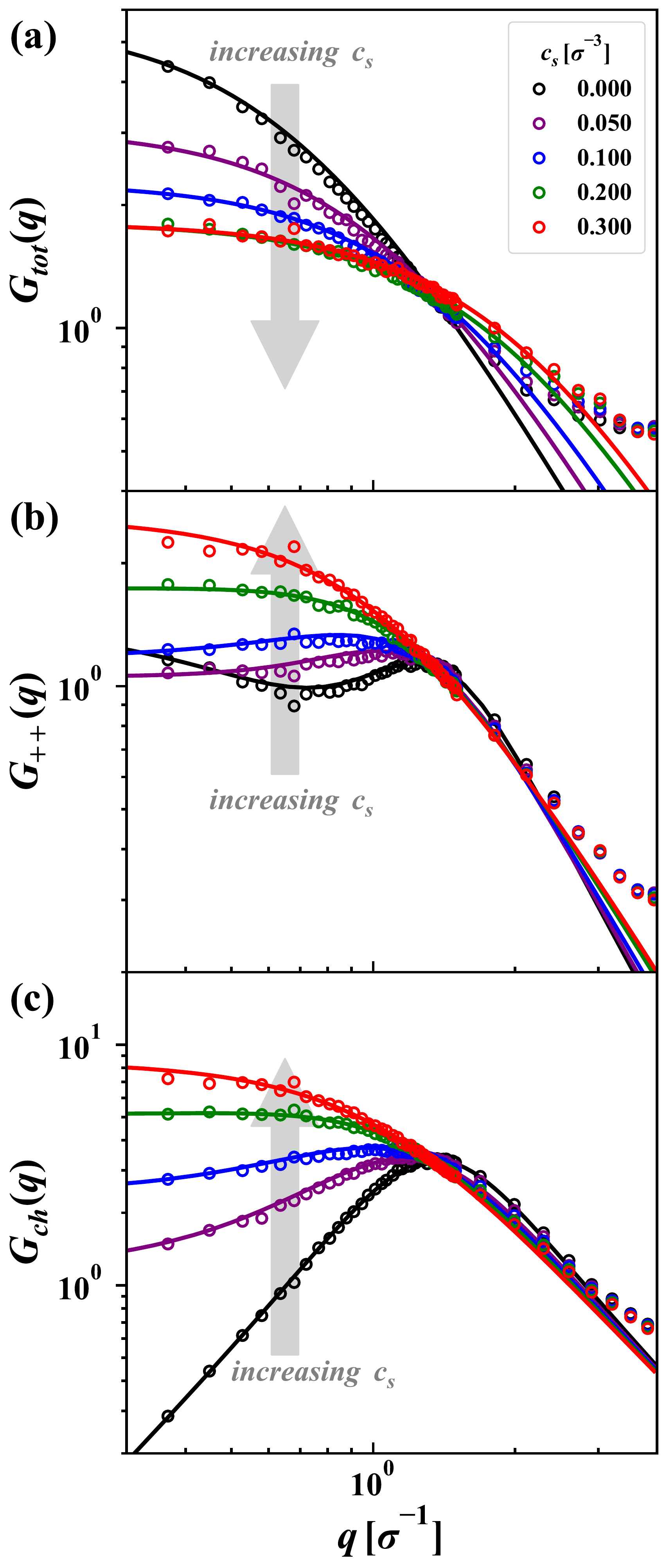}
\caption{Structure factor of salt-added PECs from simulations (open circles) and theory (solid lines):
(a) total structure factor, $G_{tot} (q)$;
(b) structure factor of polycations, $G_{++} (q)$;
(c) charge structure factor, $G_{ch} (q)$.
Simulation parameters are equal to $N = 51$, $f = 1/3$, and $l_B / \sigma = 1$. Salt concentrations are equal to $c_s \sigma^3 = 0$, $0.050$, $0.100$, $0.200$, and $0.300$. The functional form of the theoretical fitting curves are given by eqs.~\ref{gtot-fit}, \ref{relation-of-gs}, and~\ref{gch-fit}, respectively.}
\label{fig:theorscat}
\end{figure}

With screening from increasing salt concentration, the net charge of the blobs and their Coulomb interactions decrease. This diminishes the extent to which the positively charged blobs are preferentially surrounded by negatively charged blobs (Figure~\ref{fig:chart}b, blue star). At high salt concentrations, when Coulomb interactions between the polyelectrolytes are almost entirely screened (Figure~\ref{fig:chart}b, red star), each blob has an equal number of polyanion and polycation blob neighbors. Charge correlations disappear and polyelectrolytes of opposite charge become effectively indistinguishable. Polyanions and polycations form interpenetrating quasi-neutral semidilute solutions, and their structure factors $G_{++} (q) = G_{--} (q)$ adopt the OZ form. Thus, when the salt concentration is sufficiently high, the $G_{--} (q)$ and $G_{ch} (q)$ peaks, which are indicative of positional charge correlations, vanish.

The same conclusions can be drawn more rigorously via analysis of the charge structure factor, which is derived theoretically and given by eq.~\ref{g++salty-BFH}:
\begin{equation}
\label{gch-fit}
G_{ch} (Q) \propto
\frac{ 1 } { \dfrac{1}{Q^2 + s} + Q^2 + R}.   
\end{equation}
Here the dimensionless parameter $s = r_p^2 / r_D^2 \propto c_s$ describes the competition between polyelectrolytes and salt ions in screening Coulomb interactions;
The Debye radius due to salt, $r_D$, is defined by eq.~\ref{Debye-rad}. As the salt concentration increases, the peak becomes more diffuse and its position shifts to lower wavevectors, $Q^{*} = \sqrt{1-s}$. Finally, at $s \geq 1$, the peak completely disappears. The predicted behavior of the charge structure factor is observed in our simulations, shown in Figure~\ref{fig:theorscat}. Peaks in both $G_{++} (q)$ and $G_{ch} (q)$ vanish upon the addition of salt and, at high salt concentrations, all structure factors adopt the familiar OZ form, as expected within the RPA. 

\begin{figure}
\centering
\includegraphics[width=0.8\linewidth]{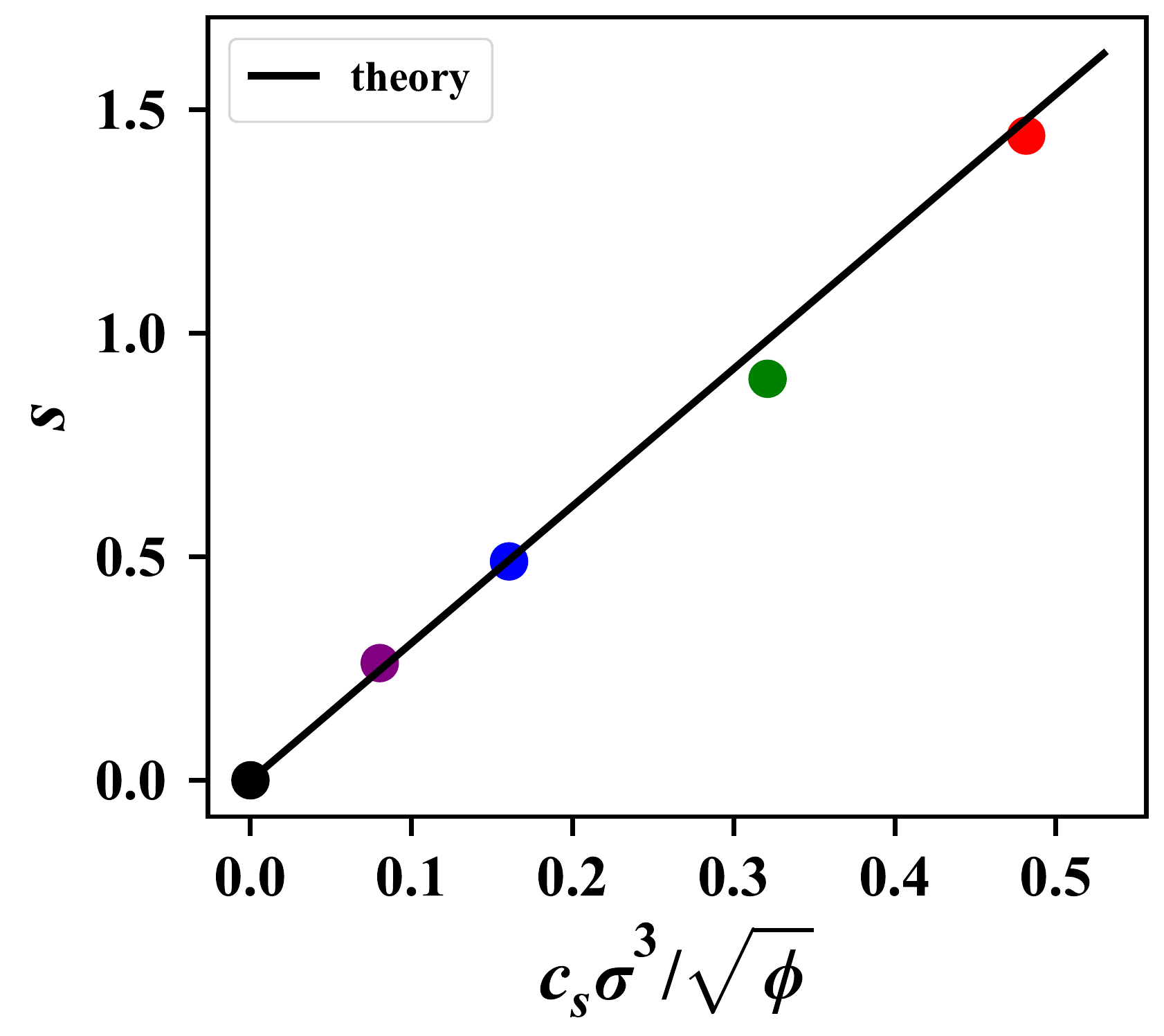}
\caption{Dependence of parameter $s$ on salt concentration $c_s$. Dots denote the values obtained from fitting the simulation data, and the solid
line shows the theoretical dependence given by eq.~\ref{s-param}. In simulations, the polymer volume fraction is calculated as $\phi = c_{p} \sigma^3$.}
\label{fig:s-vs-c_s}
\end{figure}

To demonstrate that the agreement between theory and simulations is quantitative, we simultaneously fit the scattering profiles from simulations (dots in Figure~\ref{fig:theorscat}) with the respective theoretical functional forms. Namely, we use
\begin{equation}
\label{gtot-fit}
G_{tot} (Q) \propto
\frac{1}{Q^2 + t}   
\end{equation}
for the total structure factor, eq.~\ref{gch-fit} for the charge structure factor, $G_{ch} (q)$, and their combination defined by eq.~\ref{relation-of-gs} for the structure factor of polycations. Theoretical results in Figure~\ref{fig:theorscat} are shown with solid lines. We emphasize that only three parameters---$s$, $t$, and $R$---have been used to fit the entire set of scattering profiles at any salt concentration because the $r_p$ has remained unchanged for all the curves at all $c_s$. Moreover, these parameters are microscopic rather than phenomenological, as they appear in the RPA theory and have a clear physical meaning (see RPA theory section). This is supported by the dependence of the key parameter, $s$, which governs the evolution of the scattering profiles with addition of salt, on $c_s$. That dependence is shown in Figure~\ref{fig:s-vs-c_s} and follows the theoretically expected law, eq.~\ref{s-param}.
Not only is it a linear dependence, but the numerical value of the slope coincides exactly with the theoretical prediction. In line with the RPA, the scattering peak disappears at $s$ sufficiently close to unity. The resulting negative values of $R$ (\textit{SI Appendix}, Figure S2) suggest an effective incompatibility between polyanions and polycations, which arises owing to the correlation-induced renormalization of short-range interactions between them and cannot be taken into account at the level of the RPA. The local packing effects due to short-range repulsions between the monomers and charge discreteness, which were also neglected in the theory, are another reason for that.

\begin{figure}
\centering
\includegraphics[width=0.8\linewidth]{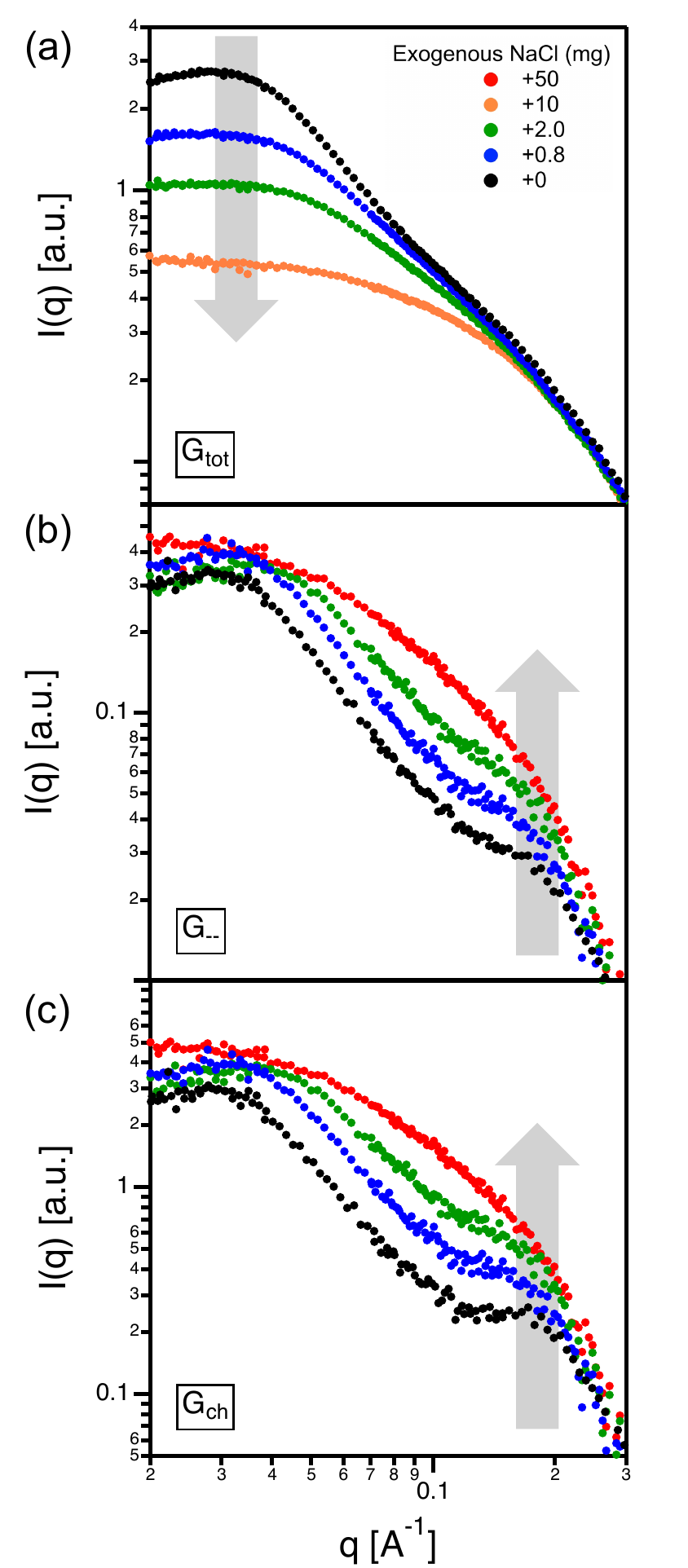}
\caption{Background subtracted scattering intensity $I (q)$ as a function of a wavevector $q$ obtained from SANS experiments on salt-added PECs with the fraction $f = 0.30$ of ionic monomers. 
(a) $G_{tot}$ is obtained via complexation of fully protonated chains of 
poly(Am$_{\text{ox}}$-{\it stat}-EO) and poly(Sulf-{\it stat}-EO) and measured against \ch{D2O};
(b) $G_{--}$ is obtained via complexation of fully protonated poly(Sulf-{\it stat}-EO) and partially deuterated poly(Am$_{\text{ox}}$-{\it stat}-$d_4$-EO)
and measured against \ch{H2O} / \ch{D2O} $= 49/51$ that contrast matches deuterated polycation;
(c) $G_{ch}$ is obtained by subtracting $G_{tot}$ from $G_{--}$ through the theoretical equation $G_{ch} = 4 G_{--} - G_{tot}$ after their appropriate normalization at high $q$ values.}
\label{fig:expscat-salt}
\end{figure}

Figure~\ref{fig:expscat-salt} shows experimental scattering results for the three structure factors. The decrease and gradual disappearance of the peaks in $G_{--} (q)$ and $G_{ch} (q)$ with addition of salt provides strong support for the view that the peaks arise from positional charge correlations. As discussed for the "salt-free" PEC data, samples prepared without exogenous salt contain $< 0.6$ mg NaCl. With addition of 0.8~mg NaCl this implies a total mass of salt $ < 1.4$~mg NaCl and with addition of 2.0 mg NaCl a total mass of NaCl $ < 2.6$ mg. The salt resistance for the samples presented here would be higher than what reported previously due to the difference in $c_{P,i}$. A faint correlation peak can still be observed with addition of 2.0~mg NaCl, which corresponds to $c_{s} \approx$ 0.74 wt \% (2.6 mg NaCl/352 mg PEC) exceeding the salt resistance previously reported, $c_{s} \approx$ 0.55 wt \%~\cite{neitzel-2021}. Under these conditions, the Debye radius $r_{D} \approx 0.8$ nm remains higher than the polymer screening radius, $r_p \approx 0.5$ nm, so that $s < 1$ and the peak is indeed expected to survive. The exogenous salt added here, in amounts of 10 and 50~mg NaCl, is well above the salt resistance and provides $s > 1$, and therefore all structure factors return to an OZ form. Above the salt resistance, a complete screening of Coulomb interactions results in a quasi-neutral semidilute solution. 

The evolution of $G_{tot} (q)$ at increasing $c_s$ is also consistent across theory, simulations, and experiments. Figures~\ref{fig:theorscat}a and \ref{fig:expscat-salt}a demonstrate that the plateau value of the OZ curve, which is equal to the osmotic compressibility of the PECs, eq.~\ref{scat-vs-osmotic}, decreases. The PEC compressibility is defined by the balance of excluded volume repulsions and Coulomb attractions. At a fixed polymer density, the addition of salt screens attractions but does not affect repulsions, leading to a decrease in PEC compressibility. The ratio between the $G_{tot} (q \to 0)$ plateau values for the salt-free and salted PECs is close to $3$ in both experiments and simulations. The RPA predicts the same trend with a lower factor of $1.3$ for $\Theta$ solvent conditions, as follows from eq.~\ref{Edw-length} (SI Appendix). This discrepancy can be primarily attributed to the applicability range of the virial expansion employed in the theory, $\phi \ll 1$, and the fact that the systems considered in our simulations are relatively dense, e.g., $\phi = c_{p} \sigma^3 \approx 0.388$. Indeed, if within the RPA the solvent is assumed to be athermal, and two-body repulsions dominate over three-body, the factor of 1.3 changes to 4 (SI Appendix). It can also be seen in Figures~\ref{fig:theorscat}a and \ref{fig:expscat-salt}a that salt extends the scattering plateau in $G_{tot} (q)$ to higher $q$ values. This is consistent with the theoretical predictions of eqs.~\ref{Edw-length}–\ref{gtot-final}, which suggest that $\xi_E$ and the parameter $t$ are decreasing functions of $c_s$ and that the plateau's right edge, $q_{E} \simeq \xi^{-1}$, increases with $c_s$.  

\subsection*{Relation to Microphase Separation Peak}

The observed peak in $G_{ch} (q)$ is analogous to the scattering peak indicative of electrostatically stabilized microphase separation in a PEC phase. The latter has been predicted for PECs/blends of weakly and oppositely charged, chemically incompatible polyelectrolytes~\cite{RKB-2018, RGK-2019, RdP-2020, fredrickson-2021, semenov-2021} and was recently observed in SAXS experiments~\cite{segalman-2022}. The $q$-values of the charge correlation peak reported here and the peak indicative of microphase separation in PEC are predicted to coincide, $q^{*} = r_p^{-1}$, which reflects the closely related physics of the two phenomena~\cite{RKB-2018}. Theory suggests that the charge correlation peak continuously grows and sharpens as $\chi_{+-}$ increases, which indicates a strong increase of positional correlations between polyanions and polycations. At the spinodal, the RPA predicts the divergence of the charge structure factor~\cite{RKB-2018}, as microphase separation is a weak crystallization phase transition~\cite{kats-1993} from a liquid system with short-range correlations (considered here) to a system with long-range order. The structure factor divergence is corrected~\cite{RdP-2020} when fluctuations are taken into account within the Brazovskii-Fredrickson-Helfand approach~\cite{brazovskii-1975, fredrickson-1987, DE-1991}. Exogenous salt suppresses charge correlations in the system considered here as well as electrostatically stabilized microphases. The RPA predicts that the peak in $G_{ch} (q)$ disappears at the same (reduced) salt concentration, $s=1$. For electrostatically stabilized microphases, this corresponds to the coordinate of the Lifshitz point where the system transitions from microphase to macrophase separation.~\cite{RKB-2018, fredrickson-2021, semenov-2021, fredrickson-2022}

\section*{Conclusions}

We have provided direct experimental evidence in support of the existence of positional charge correlations in liquid polyelectrolyte complexes. Synthetic polyelectrolytes were carefully designed and SANS measurements with contrast matching were performed to probe a long-standing theoretical picture. Oxyanionic copolymerization of allyl glycidyl ether with ethylene oxide or deuterated ethylene oxide provided well-defined neutral precursors to near-ideally random copolyelectrolytes with or without deuterium labeling and charge densities of $f \approx 0.3$. The total polymer structure factor, $G_{tot} (q)$, obtained by scattering from complexes of oppositely charged polyelectrolytes without deuterium labeling, is in good agreement with previously reported scattering profiles displaying an Orstein-Zernike form. The polyanion structure factor, $G_{--} (q)$, accessed by contrast-matching solvent and deuterium-labeled polycations, exhibited a shoulder at $q^{*} \approx 0.2~\angstrom^{-1}$. The experimental charge structure factor, $G_{ch} (q)$, was calculated by subtracting the appropriately normalized $G_{--} (q)$ from $G_{tot} (q)$. A clear peak emerged in $G_{ch} (q)$ at $q^{*}$, which is attributed to positional correlations between charged monomers of polyanions and polycations. The peak position is defined by the electrostatic screening radius from polyelectrolytes, which is found to equal $r_p  = 1 / q^{*} \approx 0.5~\text{nm}$. 

As salt was added to PECs at constant polymer density, the correlation peaks in $G_{--} (q)$ and $G_{ch} (q)$ gradually disappeared, returning all scattering profiles to a simple Ornstein-Zernike shape. The correlation peaks and their salt-induced suppression were predicted by the RPA analysis and observed in accompanying coarse-grained simulations. While field theory~\cite{BE-1988, BE-1990, joanny-2000, olvera-2004, QdP-2016, RP-2017, RKB-2018} provides a more accurate and detailed description of the positional charge correlations in PECs, they can be qualitatively interpreted using a scaling approach~\cite{remark-1}. If the PEC is viewed as a melt of oppositely charged electrostatic blobs of size $\xi \simeq \xi_{e}$, the idea of charge correlations suggests that each blob in the salt-free PEC is preferentially surrounded by oppositely charged blobs~\cite{SZDR-2005, RZB-2018, rubinstein-2018}. These local positional correlations result in the peak in the charge structure factor, $G_{ch} (q)$, and, hence, in the polyanion structure factor, $G_{--} (q)$, at $q^{*} \simeq \xi^{-1} \simeq r_p^{-1}$. The estimate of the blob size as the half-period of oscillations in $G_{ch} (r)$ yields $\xi \approx D/2 = \sqrt{2} \pi r_p \approx 2.2$~nm. Addition of salt results in the screening of Coulomb interactions, thereby diminishing the preferential adjacency of oppositely charged blobs until the PEC structurally resembles a mixed semidilute solution of barely distinguishable, quasi-neutral polyanions and polycations. 

Collectively, our combined results from theory, simulations, and experiments make a strong case for the existence of positional charge correlations as a fundamental feature of polyelectrolyte complexes.

\section*{Acknowledgement}
This work was supported by the Department of Energy, Basic Energy Sciences, Division of Materials Science and Engineering. A portion of this research used resources at the Spallation Neutron Source, a DOE Office of Science User Facility operated by the Oak Ridge National Laboratory.  We gratefully acknowledge Carrie Gao for her assistance during the neutron scattering experiments.


\clearpage

\end{document}


\title{Supporting Information for \\
Scattering Evidence of Positional Charge Correlations in Polyelectrolyte Complexes}

\author{Yan Fang$^{a,b}$}
\thanks{equal contribution}
\author{Artem M. Rumyantsev$^{a,c}$}
\thanks{equal contribution}
\author{Angelika E. Neitzel$^{a,b,d}$}
\author{Heyi Liang$^{a}$}
\author{William T. Heller$^{e}$}
\author{Paul F. Nealey$^{a,b}$}
\author{Matthew V. Tirrell$^{a,b}$}
\email{mtirrell@uchicago.edu}
\author{Juan J. de Pablo$^{a,b}$}
\email{depablo@uchicago.edu}

\affiliation{
 $^a$Pritzker School of Molecular Engineering, University of Chicago, Chicago, Illinois 60637, United States \\
 $^b$Center for Molecular Engineering and Materials Science Division, Argonne National Laboratory, Lemont, Illinois 60439, United States \\
 $^c$Department of Chemical and Biomolecular Engineering, North Carolina State University, Raleigh, North Carolina 27695, United States \\
 $^d$Department of Materials Science \& Engineering, University of Florida, Gainesville, FL 32611 \\
 $^e$Neutron Scattering Division, Oak Ridge National Laboratory, Oak Ridge, Tennessee 37831, United States
}

\date{\today}

\maketitle


\section{RPA Theory of Scattering}

\subsection{Inversion of the Structure Factor Matrix in the Salt-Free Case}

The elements of the inverse structure factor matrix within the RPA are given by eqs. 6-7 of the manuscript: 
\begin{equation}
G_{++}^{-1} = G_{--}^{-1} = \frac{q^2}{6 \phi} + \frac{4 \pi u f^2}{q^2} + 6 w \phi
\end{equation}
\begin{equation}
G_{+-}^{-1} = G_{-+}^{-1} = - \frac{4 \pi u f^2}{q^2} + 6 w \phi
\end{equation}
This matrix can be written as 
\begin{equation}
G^{-1} = 
\begin{pmatrix}
\dfrac{q^2}{A} + \dfrac{B}{q^2} + C & - \dfrac{B}{q^2} + C \\
- \dfrac{B}{q^2} + C & \dfrac{q^2}{A} + \dfrac{B}{q^2} + C
\end{pmatrix}
\end{equation}
with $A = 6 \phi$, $B = 4 \pi u f^2$, and $C = 6 w \phi$. It should be inverted. The determinant is given by
\begin{equation}
\det G^{-1} = 
\left( \dfrac{q^2}{A} + \dfrac{2B}{q^2} \right) 
\left( \dfrac{q^2}{A} + 2C  \right)
\end{equation}
and the inverse matrix reads
\begin{equation}
G = \dfrac{1}{ \det G^{-1} }
\begin{pmatrix}
\dfrac{q^2}{A} + \dfrac{B}{q^2} +C & \dfrac{B}{q^2} -C \\
\dfrac{B}{q^2} - C & \dfrac{q^2}{A} + \dfrac{B}{q^2} + C
\end{pmatrix}
\end{equation}
The total structure factor is given by
\begin{multline}
G_{tot} (q) = 2 \left( G_{++} + G_{+-} \right) = 2 \; 
\dfrac{ \dfrac{q^2}{A} + \dfrac{2B}{q^2} }{ \left( \dfrac{q^2}{A} + \dfrac{2B}{q^2} \right) \left( \dfrac{q^2}{A} + 2C  \right) } = 
\frac{ C^{-1} }{ 1 + \dfrac{q^2}{2AC} } = \\
\frac{ \left( 6 w \phi \right)^{-1} }{ 1 + \left( q \xi_{E} \right)^{2} }
= \sqrt{ \dfrac{3 \phi}{\pi u f^2} } \cdot \frac{1}{Q^2 + t}
\end{multline}
Here the Edwards screening length is defined by $\xi_{E}^{-2} = 2 A C = 72 w \phi^2 $, $t = r_p^2 / \xi_E^2$, and the polymer screening radius of Coulomb interactions is defined by $r_p^{-4}  = 2 A B  = 48 \pi u f^2 \phi$.~\cite{BE-1988}
Similarly, polymer charge structure factor reads
\begin{multline}
G_{ch} (q) = 2 \left( G_{++} - G_{+-} \right) = 
2 \; \dfrac{ \dfrac{q^2}{A} + 2C } { \left( \dfrac{q^2}{A} + \dfrac{2B}{q^2} \right) \left( \dfrac{q^2}{A} + 2C  \right) } = 
\dfrac{  \dfrac{q^2}{B} }{ 1 +  \dfrac{q^4}{2AB} } = \\
\dfrac{1}{B r_p^2} \cdot
\dfrac{ \left( q r_p \right)^2 } { 1 + \left( q r_p \right)^4 }
= \sqrt{ \dfrac{3 \phi}{\pi u f^2} } \cdot \dfrac{Q^2}{ 1 + Q^4}
\label{G_ch-final-theta}
\end{multline}

\subsection{Inversion of the Structure Factor Matrix in the Salt-Added Case}

The elements of the inverse structure factor matrix within the RPA in the presence of salt are given by
\begin{equation}
G_{++}^{-1} = G_{--}^{-1} = \frac{q^2}{6 \phi} + \frac{4 \pi u f^2}{q^2 + r_D^{-2} } + 6 w \phi
\end{equation}
\begin{equation}
G_{+-}^{-1} = G_{-+}^{-1} = - \frac{4 \pi u f^2}{q^2 + r_D^{-2} } + 6 w \phi
\end{equation}
where $r_D = \left( 4 \pi u c_s \right)^{-1/2}$ is the Debye radius due to small salt ions. This matrix can be written as 
\begin{equation}
G^{-1} = 
\begin{pmatrix}
\dfrac{q^2}{A} + \dfrac{B}{q^2 + r_D^{-2}} + C & - \dfrac{B}{q^2 + r_D^{-2}} + C \\
- \dfrac{B}{q^2 + r_D^{-2}} + C & \dfrac{q^2}{A} + \dfrac{B}{q^2 + r_D^{-2}} + C
\end{pmatrix}
\end{equation}
where the definition of the introduced variables remains unchanged, $A = 6 \phi$, $B = 4 \pi u f^2$, and $C = 6 w \phi$. To invert this matrix, we first calculate its determinant:
\begin{equation}
\det G^{-1} = 
\left( \dfrac{q^2}{A} + \dfrac{2B}{q^2 + r_D^{-2}} \right) 
\left( \dfrac{q^2}{A} + 2C  \right)
\end{equation}
The inverse matrix equals
\begin{equation}
G = \dfrac{1}{ \det G^{-1} }
\begin{pmatrix}
\dfrac{q^2}{A} + \dfrac{B}{q^2 + r_D^{-2}} +C & \dfrac{B}{q^2 + r_D^{-2}} -C \\
\dfrac{B}{q^2 + r_D^{-2}} - C & \dfrac{q^2}{A} + \dfrac{B}{q^2 + r_D^{-2}} + C
\end{pmatrix}
\end{equation}
The total structure factor
\begin{multline}
G_{tot} (q) = 2 \left( G_{++} + G_{+-} \right) = 2 \; 
\dfrac{ \dfrac{q^2}{A} + \dfrac{2B}{q^2 + r_D^{-2}} }{ \left( \dfrac{q^2}{A} + \dfrac{2B}{q^2 + r_D^{-2}} \right) \left( \dfrac{q^2}{A} + 2C  \right) } = 
\frac{ C^{-1} }{ 1 + \dfrac{q^2}{2AC} } = \\
\frac{ \left( 6 w \phi \right)^{-1} }{ 1 + \left( q \xi_{E} \right)^{2} }
 = \sqrt{ \dfrac{3 \phi}{\pi u f^2} } \cdot \frac{1}{Q^2 + t}
\end{multline}
Here the Edwards screening length and the reduced solvent quality $t$ remain unchanged. The charge structure factor changes and, in the presence of salt, is given by
\begin{multline}
G_{ch} (q) = 2 \left( G_{++} - G_{+-} \right) = 
2 \; \dfrac{ \dfrac{q^2}{A} + 2C } { \left( \dfrac{q^2}{A} + \dfrac{2B}{q^2 + r_D^{-2}} \right) \left( \dfrac{q^2}{A} + 2C  \right) } = \\
\frac{ \sqrt{\dfrac{2A}{B} } }
{ \dfrac{q^2}{\sqrt{2AB}} + \dfrac{ \sqrt{2 A B} } {q^2 + r_D^{-2}} }
= \sqrt{ \dfrac{3 \phi}{\pi u f^2} } \cdot \dfrac{1}{ \dfrac{1}{Q^2 + s} + Q^2}
\end{multline}
Here we made use of $\sqrt{2 A B} = r_p^{-2}$ and introduced $s = r_p^2 / r_D^2$.~\cite{BE-1988} Finally, the structure factor of polyanions (and polycations, owing to the system symmetry) can be written as
\begin{multline}
G_{++} (q) = G_{--} (q) = \frac{G_{tot} (q) + G_{ch} (q) }{4} = 
\frac{1}{4}\sqrt{ \dfrac{3 \phi}{\pi u f^2} } \cdot \left[ \frac{1}{Q^2 + t} + \frac{1} {\dfrac{1}{Q^2 + s} + Q^2 } \right] = \\
\frac{1}{4}\sqrt{ \dfrac{3 \phi}{\pi u f^2} } \cdot \dfrac{2 Q^4 + Q^2 (2 s + t) + st + 1}
{\left( Q^2 + t \right) \left( 1 + Q^2 s + Q^4 \right)}
\end{multline}
We note that a similar RPA-based approach~\cite{BE-1988, JL-1990} has been earlier used for predicting microphase separation in polyelectrolyte solutions/gels under poor solvent conditions and for interpreting the respective experimental scattering profiles.~\cite{exp-1990, exp-1992}

\subsection{Total Structure Factor and Osmotic Compressibility}

The equilibrium coacervate density is given by the equality of the osmotic pressures,
$\Pi = \phi \cdot {d F} / {d \phi} - F$, corresponding to the three body repulsive and correlation attractive parts of the free energy. In $\Theta$ solvent, $F_{vol} = w \phi^3$ and $\Pi_{vol} = 2 w \phi^3$. For the attractive part, one should use
\begin{equation}
F_{corr} = (1-s) \sqrt{2+s} \; \frac{ \left( 48 \pi u f^2 \right)^{3/4} }{ 6 \sqrt{2} \pi } \phi^{3/4}
\end{equation}
and 
\begin{equation}
\Pi_{corr} = \frac{\partial F_{corr}}{ \partial \phi} +     
\frac{\partial F_{corr}}{ \partial s}  \frac{d \phi}{d s}  
\end{equation}
In the salt-free case, $c_s = 0$ and $s = 0$, the correlation free energy and pressure take simple forms:
\begin{equation}
F_{corr} = \frac{ \left( 48 \pi u f^2 \right)^{3/4} }{ 6 \sqrt{2} \pi } \phi^{3/4}; 
\qquad  
\Pi_{corr} = - \frac{1}{4} \cdot \frac{ \left( 48 \pi u f^2 \right)^{3/4} }{ 6 \sqrt{2} \pi } \phi^{3/4}
\end{equation}
The balance between the pressures defines the equilibrium coacervate density
\begin{equation}
\phi_{0} = \frac{1}{2^{2/3} \left( 3 \pi \right)^{1/9}} u^{1/3} f^{2/3} w^{-4/9} \approx 0.49 u^{1/3} f^{2/3} w^{-4/9} 
\end{equation}
We can now explicitly calculate the Edwards correlation length 
\begin{multline}
\xi_{E}^{-2} = 12 \phi_{0} \left( 6 w \phi_{0} - 
\dfrac{3}{16} \cdot \frac{ \left( 48 \pi u f^2 \right)^{3/4} }{ 6 \sqrt{2} \pi } \phi_{0}^{-5/4} \right) = \\
12 \phi_{0} \left( 2.94 \cdot u^{1/3} f^{2/3} w^{5/9} - 0.74 \cdot u^{1/3} f^{2/3} w^{5/9} \right) = 12 \phi_{0} \cdot 2.2 u^{1/3} f^{2/3} w^{5/9} 
\end{multline}
Recall that the osmotic compressibility of the coacervate is given by
\begin{equation}
G_{tot} (q \to 0)
= \left[ \frac{d \left( \Pi_{vol} + \Pi_{corr} \right)} {d \phi} \right]^{-1} 
= \left[ \phi \frac{d^2 \left( F_{vol} + F_{corr} \right) } {d^2 \phi} \right]^{-1} \end{equation}
For salt-free $\Theta$ solvent, the total structure factor of the system equals
\begin{equation}
\left. G_{tot} (q) \right|_{c_s = 0} = \frac{12 \phi_{0}}{ q^2 + \xi_{E}^{-2} } \to 
\frac{1} { 2.2 u^{1/3} f^{2/3} w^{5/9} } = 0.45 u^{-1/3} f^{-2/3} w^{-5/9}
\end{equation}
at $q \to 0$. If the density of the coacervate, $\phi_{0}$, remains unchanged but a lot of salt is added to the system, $F_{corr} \to 0$ and   
\begin{equation}
\xi_{E}^{-2} = 12 \phi_{0} \cdot \left( 6 w \phi_{0} \right) = 12 \phi_{0} \cdot 2.94 u^{1/3} f^{2/3} w^{5/9} 
\end{equation}
This results in the structure factor asymptotic behavior
\begin{equation}
\left. G_{tot} (q) \right|_{c_s \to \infty} = \frac{12 \phi_{0}}{ q^2 + \xi_{E}^{-2} } \to 
\frac{1} { 2.94 u^{1/3} f^{2/3} w^{5/9} } = 0.34 u^{-1/3} f^{-2/3} w^{-5/9}
\end{equation}
at $q \to 0$. Thus, the addition of salt should result in the shifting $G_{tot} (q=0)$ down by approximately 25\% as compared to the salt-free case, $0.45 / 0.34 = 1.32$. 

We note that the decrease of $G_{tot} (q=0)$ is {\it very sensitive to the solvent quality}. Consider good solvent where two-body repulsions dominate over three-body interactions. For simplicity, assume that excluded volume interactions are described by $F_{vol} = v \phi^2$ while the correlation correction coincides with that obtained for $\Theta$ solvent, 
$F_{corr} = A \phi^{3/4}$. (In fact, in the athermal solvent, $F_{vol} \sim \phi^{9/4}$~\cite{degennes-book}; for the correlation term, more refined calculations using swollen coil sturture factor~\cite{BE-1990} suggest that $ F_{corr} \sim \phi^{9/11}$). In this case 
\begin{equation}
\phi_{0} = \left( \frac {A} {4 v} \right)^{4/5}
\end{equation}
\begin{equation}
\frac {d^2 F_{vol}} {d \phi^2} = v
\end{equation}
\begin{equation}
\frac{d^2 F_{corr}} {d \phi^2} = - \frac{3}{16} A \phi_{0}^{-5/4} = - \frac{3 v}{4}
\end{equation}
These estimates show that the osmotic compressibility drop in the athermal is much higher as compared to $\Theta$ solvent:
\begin{equation}
\frac{ \left. G_{tot} (q = 0) \right|_{c_s = 0} } {\left. G_{tot} (q = 0) \right|_{c_s \to \infty}} =
= \dfrac{ \dfrac {d^2 F_{vol}} {d \phi^2}} 
{ \dfrac {d^2 F_{vol}} {d \phi^2} + \dfrac {d^2 F_{corr}} {d \phi^2} }
= \frac{v}{ \left( v - 3 v / 4 \right) } = 4
\end{equation}
This numerical factor equal to $4$ better corresponds to the experimentally observed value of $3$, suggesting that the solvent quality for the polyelectrolytes is good and that they are locally swollen. This is also consistent with the slope of $-1.6 \approx 1/ \nu_{+}$ in the total structure factor, with $\nu_{+} = 0.588$ being the swollen coil exponent.

\subsection{Charge Structure Factor and Correlation Peak in the Athermal Solvent}

The RPA analysis in the manuscript's main text was devoted to the case of $\Theta$ solvent when PE chains have Gaussian conformations at all length scales. For the athermal solvent, this is not the case because PE chains swell at small length scales, within the concentration blobs. The salt-free coacervate should be viewed as a conglomerate of swollen electrostatic blobs, with self-avoiding random walk statistics inside them.~\cite{BE-1988, BE-1990, rubinstein-2006, RZB-2018} At the same time, the PE considered as a chain of the swollen blobs exhibits Gaussian statistics, i.e., polyions retain ideal-coil conformations at the length exceeding the concentration blob size.~\cite{RZB-2018} This is in complete analogy with the neutral semidilute solutions in the athermal solvents.~\cite{degennes-book} Therefore, the structure factor of PEs has a more complicated form, which does not enable deriving analytics results, which were available for $\Theta$ solvent.

However, the approximation for the structure factor can be made to demonstrate the existence of the correlation peak in the athermal solvent (subscript ``a''). Screening of Coulomb interactions depends on the PE conformational statistics at the lowest lengths, within the blobs, and is almost independent of that at larger length scales.~\cite{BE-1988, RZB-2018} Therefore, the PE structure factor is approximated by 
\begin{equation}
G_{0,a} = \frac{\alpha \phi} { q^{5/3} }
\end{equation}
where $\alpha \simeq 1$ is a (unknown) numerical constant on the order of unity. Here we have used the Flory value of the critical exponent, $\nu = 5/3$. This structure factor properly describes locally swollen PE statistics. We note that the contributions due to non-Coulomb interactions are absent because the effect of the solvent quality is already taken into account by the appropriate choice of $G_{0,a} (q) \sim q^{-5/3}$. For salt-free coacervates, the inverse elements of the structure factor matrix take the following form:
\begin{equation}
G_{++}^{-1} = G_{--}^{-1} = \frac{q^{5/3} }{\alpha \phi} + \frac{4 \pi u f^2}{q^2}
\end{equation}
\begin{equation}
G_{+-}^{-1} = G_{-+}^{-1} = - \frac{4 \pi u f^2}{q^2}
\end{equation}
In the athermal solvent, the polymer screening radius of Coulomb interactions should be defined as~\cite{BE-1988, RZB-2018}
\begin{equation}
r_{p,a} = \left( 8 \pi \alpha u f^2 \phi \right)^{-3/11} \simeq 
\left(  u f^2 \phi  \right)^{-3/11}
\end{equation}
The dimensionless wavevector is hence equal to $Q = q r_{p,a}$. Inversion of the structure factor matrix yields
\begin{equation}
G_{ch,a} (q) = 2 \left( G_{++} - G_{+-} \right) = 
2 \alpha \phi \cdot \dfrac{ q^2 } { q^{11/3}  + r_{p,a}^{-11/3}}
= 2 \alpha \phi r_{p,a}^{5/3} \cdot \dfrac{Q^2}{ 1 + Q^{11/3}}
\label{G_ch-final-athermal}
\end{equation}
This functional form of the charge structure factor in the athermal solvent demonstrates that the correlation peak exists. The peak position is given by $Q^{*} = 1.05$, i.e., $q^{*} = 1.05 r_{p,a}^{-1} \simeq r_{p,a}^{-1}$. Except for the exact functional form of the structural factor (cf. eqs.~\ref{G_ch-final-theta} and \ref{G_ch-final-athermal}) and the other definition of the polymer screening radius, this result for the athermal solvent is in complete analogy with the case of $\Theta$ solvent.  

In the presence of small salt ions, the inverse structure factor matrix elements are given by
\begin{equation}
G_{++}^{-1} = G_{--}^{-1} = \frac{q^{5/3} }{\alpha \phi} + \frac{4 \pi u f^2}{q^2 + r_D^2}
\end{equation}
\begin{equation}
G_{+-}^{-1} = G_{-+}^{-1} = - \frac{4 \pi u f^2}{q^2 + r_D^2}
\end{equation}
Here the Debye radius is independent of the solvent quality; its definition coincides with that introduced earlier for $\Theta$ solvent. By inverting $G_{ij}^{-1} (q)$ matrix one can find the charge structure factor is the presence of salt:
\begin{equation}
G_{ch,a} (q) = 2 \left( G_{++} - G_{+-} \right) = 
2 \alpha \phi \cdot \dfrac{ q^2 } { q^{11/3}  + \dfrac {r_{p,a}^{-11/3}} {q^2 + r_D^{-2} }}
= 2 \alpha \phi r_{p,a}^{5/3} \cdot \dfrac{1}{ Q^{5/3} + 
\dfrac{1}{Q^2 + s_a}}
\end{equation}
Here the reduced salt concentration
\begin{equation}
s_{a} = \frac{ r_{p,a}^2 } {r_{D}^2} = \frac{4 \pi u c_s } { \left( 8 \pi \alpha u f^2 \phi \right)^{6/11} }
\sim c_s \phi^{-6/11}
\label{s_a}
\end{equation}
remain the linear function of the actual concentration of salt, $c_s$. It is the value of $s_a$ that governs the position and the very existence of the correlation peak. $s_a \ll 1$ corresponds to the limit of low salt, when the correlation peak exists and its position is given by $Q^{*} \approx 1$. As salt is added to the system and $s_a$ increases, the peak position shifts to the lower values of $Q^{*}$. It completely disappears at $s_a = 0.81$. The limiting case of high salt concentrations corresponds to $s_a \gg 1$. The qualitative evolution of the charge structure factor in the athermal solvent, $G_{ch,a} (q)$, is completely analogous to that in $\Theta$ solvent.

It should be finally noted that the inverse structure factors obtained for the athermal solvent within the RPA, $G_{ij}^{-1} (q)$ with $i,j  = +,-$,  enable correct calculations of the correlation correction to the free energy:~\cite{BE-1988}
\begin{equation}
F_{RPA} =  \frac{1}{2} \int_{0}^{q_{0}} \frac{d^3 q}{ \left( 2 \pi \right)^{3} } \ln \left( \frac{ \det \left( G_{ij}^{-1} (q) \right) }
{ \det \left. \left( G_{ij}^{-1} (q) \right) \right|_{u=0} } \right)
\end{equation}
Unfortunately, for the general salt-added case, $s_a \neq 0$, the integral cannot be calculated analytically. However, for the salt-free case this can be done to obtain $F_{RPA, a} \simeq r_{p, a}^{-3} \simeq \left( u f^2 \phi \right)^{9/11}$.~\cite{BE-1988, QdP-2016, RZB-2018} By combining this result with the scaling free energy of two-body repulsions in the athermal semidilute solutions, $F_{vol} \simeq \phi^{9/4}$,~\cite{degennes-book} one can find the equilibrium density of the salt-free athermal coacervate, $\phi \simeq \left( u f^2 \right)^{4/7}$.~\cite{RZB-2018} It coincides with the polymer density within the electrostatic blob for the athermal solvent and hence is in agreement with the scaling representation of coacervate in the athermal solvent as a densely packed array of the swollen electrostatic blobs.~\cite{rubinstein-2006, RZB-2018} This agreement demonstrates the validity of the performed RPA analysis for the athermal solvent, particularly the adopted approximation for the PE structure factor, $G_{0,a} \sim q^{-5/3}$, and the idea that screening (and hence correlation attractions) are governed by the local PE statistics within the concentration blob. This supports our theoretical conclusion on the existence and salt-induced decay of the correlation peak in the athermal solvent; it is also consistent with the results of the performed simulations shown in subsection 2\ref{sec:sim-athermal}.

\newpage
\section{Simulation Details and Results}
\label{sec:sim}

\subsection{Simulation Details}
Coarse-grained simulations of polyelectrolyte coacervates under $\Theta$ solvent conditions are performed using an implicit solvent. Polycations and polyanions are represented by the bead-spring model and both consist of $N = 51$ beads. Non-bonded beads interact with the truncated and shifted Lennard-Jones (LJ) potential
\begin{equation}
U_{LJ}(r) = 
\begin{cases}
\!\begin{aligned}
4\varepsilon_{LJ} \Bigl[
& \left(\frac{\sigma}{r}\right)^{12} - \left(\frac{\sigma}{r}\right)^{6} \\
& - \left(\frac{\sigma}{r_{cut}}\right)^{12} + \left(\frac{\sigma}{r_{cut}}\right)^{6} \Bigr], 
\end{aligned} & r < r_{cut} \\
0, & r \geq r_{cut}
\end{cases}
\label{LJ-potential}
\end{equation}
Here $r$ is the distance between the beads, $\sigma$ is the bead size, $\varepsilon_{LJ}$ is the LJ interaction parameter, and $r_{cut} = 2.5 \sigma$ is the cutoff distance. To simulate $\Theta$ solvent conditions, the LJ interaction parameter is set to $\varepsilon_{LJ} = 0.33 k_{B}T$.~\cite{grest-1993} To simulate good solvent condition, the LJ interaction parameter is set to $\varepsilon_{LJ} = 0.1 k_{B}T$. To simulate athermal solvent condition, the cutoff distance is set to $r_{cut} = 2^{1/6} \sigma$ so the LJ potential becomes purely repulsive, and the LJ interaction parameter is set to $\varepsilon_{LJ} = 1.0 k_{B}T$. The fraction of beads on each polyelectrolyte chain that are charged is $f = 1/3$ , with charge valence $z = \pm 1$. Charged beads are evenly distributed along the chain. The Coulomb interactions between charged beads are defined by
\begin{equation}
U_{coul}(r_{ij}) = k_{B}T \frac{l_{B} z_{i} z_{j}}{r_{ij}}
\label{Coul-potential}
\end{equation}
where $r_{ij}$ is the distance between beads with charge valence $z_{i}$ and $z_{j}$. The strength of the Coulomb interaction is characterized by the Bjerrum length, which is set to $l_{B} = e^2/\epsilon k_{B}T = 1.0 \sigma$, where $e$ is the elementary charge and $\epsilon$ is the dielectric constant. The Coulomb potential is evaluated by the particle-particle particle-mesh (PPPM) method~\cite{PPPM} with an estimated accuracy of $10^{-4}$. Polymer beads are connected into chains through finite-extensible nonlinear elastic (FENE) bonds. The attractive part of the FENE potential is
\begin{equation}
U_{FENE}(r) = \frac{1}{2} k_{spring} R_{max}^2 \ln
\left( 1 - \frac{r^2}{R_{max}^2} \right)
\label{FENE-potential}
\end{equation}
where $k_{spring} = 30 k_{B}T/\sigma^2$ is the spring constant, and $R_{max} = 1.5 \sigma$ is the maximum bond length. The repulsive part of the FENE potential is described by a truncated and shifted LJ potential, eq.~\ref{LJ-potential}, with $\varepsilon_{LJ} = 1.0 k_{B}T$ and $r_{cut} = 2^{1/6} \sigma$. All beads have the same mass, $m$. In all simulations, the equations of motion are updated by a velocity-Verlet algorithm with an integration timestep $\Delta t = 0.005 \tau_{LJ}$, where $\tau_{LJ} = \sigma (m/k_{B}T)^{1/2}$ is the standard LJ time. To generate the equilibrium salt-free coacervate, the simulation box is coupled to a Nos\'e-Hoover barostat with pressure $P = 0$ representing the zero osmotic pressure of polyelectrolytes in the coacervates. A constant temperature is maintained by a Nos\'e-Hoover thermostat.~\cite{LAMMPS} The equilibrium polyelectrolyte concentration of the salt-free coacervate is $c_{p} = c_{p,+} + c_{p,-} = 0.388 \sigma^{-3}$ in $\theta$ solvent, $0.100 \sigma^{-3}$ in good solvent, and $0.054 \sigma^{-3}$ in athermal solvent. To obtain the coacervate samples for the structure factor calculation, we performed simulations of coacervates with different salt concentrations in the NVT ensemble. For all coacervates considered here, the polyelectrolyte concentrations were kept constant and equal to that of the salt-free one, and charged beads were added to represent salt ions. The salt ion beads, with $z = \pm1$, interact with each other and polyelectrolyte beads as described by eqs.~\ref{LJ-potential} and~\ref{Coul-potential}. The salt concentration, $c_{s} = c_{s,+} + c_{s,-}$, is varied between $0$ and $0.3 \sigma^{-3}$ for $\theta$ solvent, and between $0$ and $0.08 \sigma^{-3}$ for good and athermal solvent. The constant temperature is in this case maintained by a Langevin thermostat. All simulations are performed using the software package LAMMPS~\cite{LAMMPS}.

\newpage

\subsection{Monomer Peak in Simulations Scattering Profiles}

\begin{figure}[h!]
\centering
\includegraphics[width=0.45\linewidth]{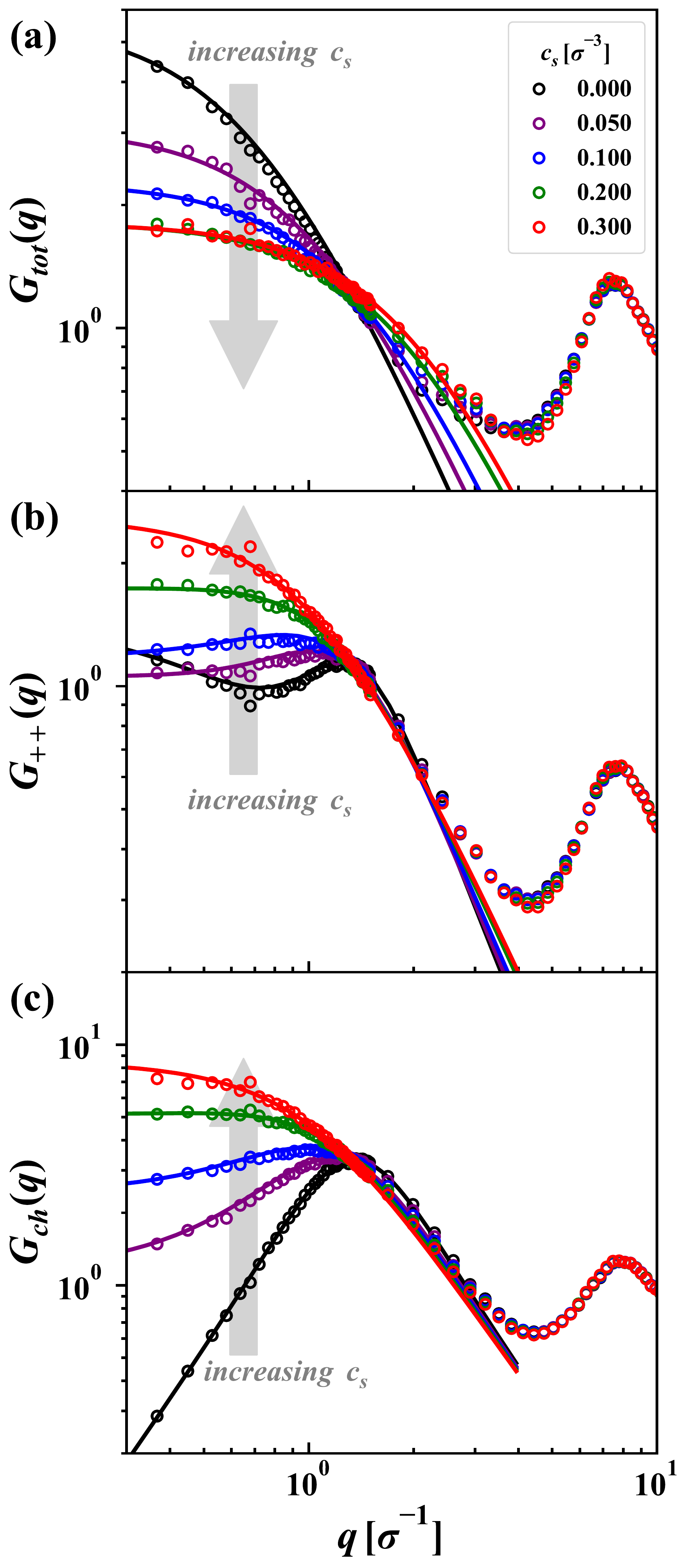}
\caption{Structure factor of coacervates under the constant polymer concentration with varying salt concentrations: 
(a) total structure factor, $G_{tot} (q)$; 
(b) polycation structure factor, $G_{++} (q)$; 
(c) charge structure factor, $G_{ch} (q)$; 
Solid lines are the best theoretical fit, with the functional form given by eqs. 30-32 of the manuscript: $G_{tot} (Q) \propto \left( Q^2 + t \right)^{-1} $, 
$G_{++} (Q) = ( G_{tot} (Q) + G_{ch} (Q) ) / 4$, and 
$G_{ch} (Q) \propto \left( 1/(Q^2 + s) + Q^2 + R \right)^{-1}$.
}
\label{fig:monomer-peak}
\end{figure}

Figure~\ref{fig:monomer-peak} is identical to Figure 5 of the main manuscript text but shows the structure factors in a wider range of $q$ to demonstrate the monomer peak.

\begin{figure}
\centering
\includegraphics[height=3.5in]{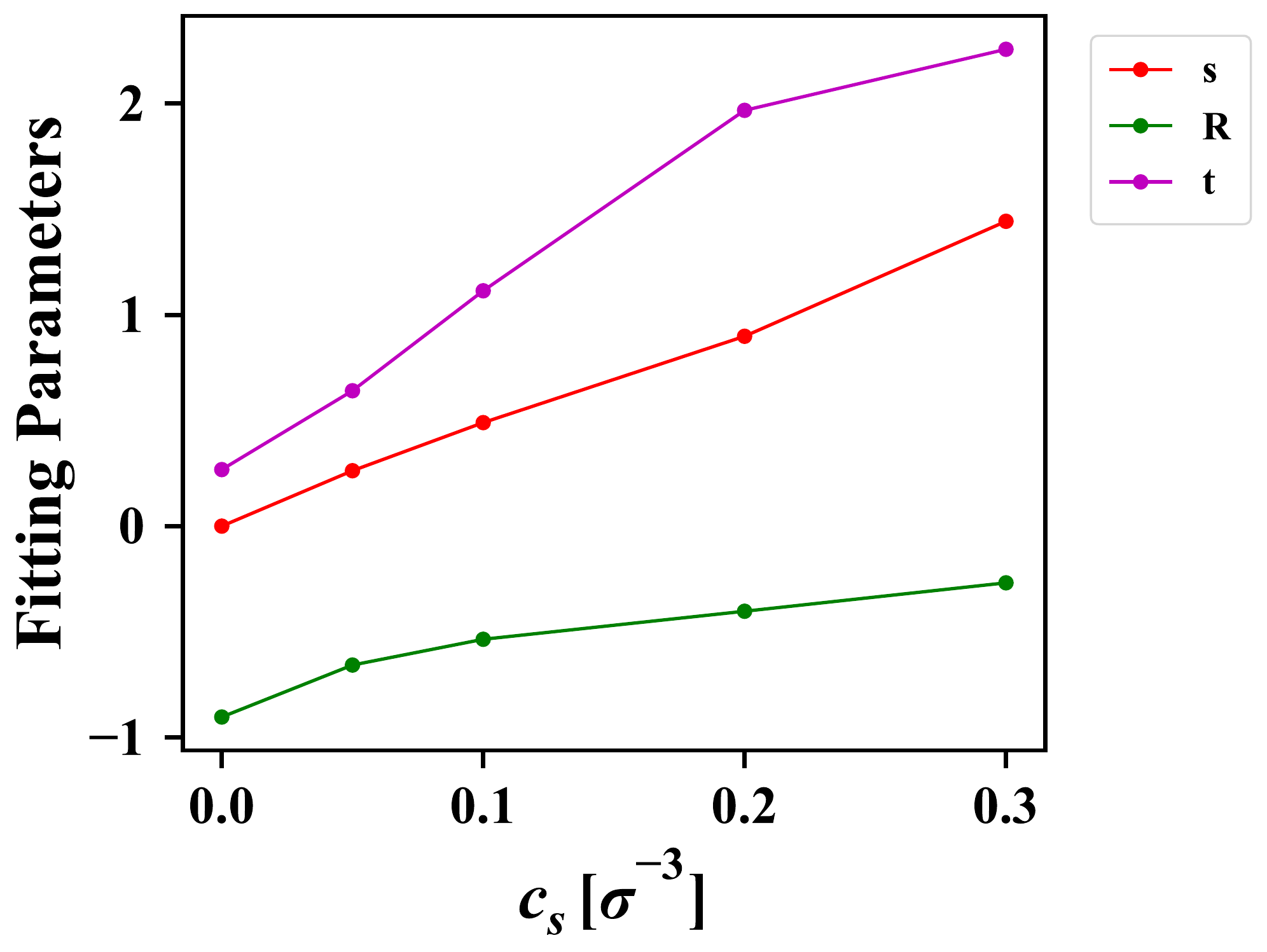}
\caption{ Dependence of the theoretical parameters $s$, $t$, and $R$, which were used to fit the simulation scattering profiles $G_{tot} (q)$, $G_{++} (q)$, and $G_{ch} (q)$ in Figure 5 of the main text, on the concentration of salt $c_s$. The reduced salt concentration $s$ increases linearly with $c_s$, as shown in Figure 6 of the main text. Negative values of $R$ should be attributed to the local packing constraints, which manifest as effective short-range incompatibility between the chains, $\chi_{+-} < 0$. The increase in the parameter $t$ is consistent with the theoretical equation 21 of the main text, which suggests that the Edwards correlation length decreases at increasing $c_s$. The latter also results in the decreasing osmotic compressibility of the coacervate, $G_{tot} ( q \to 0) \propto t^{-1}$, which is observed in simulations and the experiment.  }
\end{figure}

\newpage

\subsection{Simulation Scattering Profiles for Coacervates in the Athermal and Good Solvent}
\label{sec:sim-athermal}

The solvent quality controls the $G_{ch} (q)$ slope in the range of high $q$, which reflects the local ideal/swollen coil statistics of the polyelectrolytes.

\begin{figure} [h!]
\centering
\includegraphics[height=3.5in]{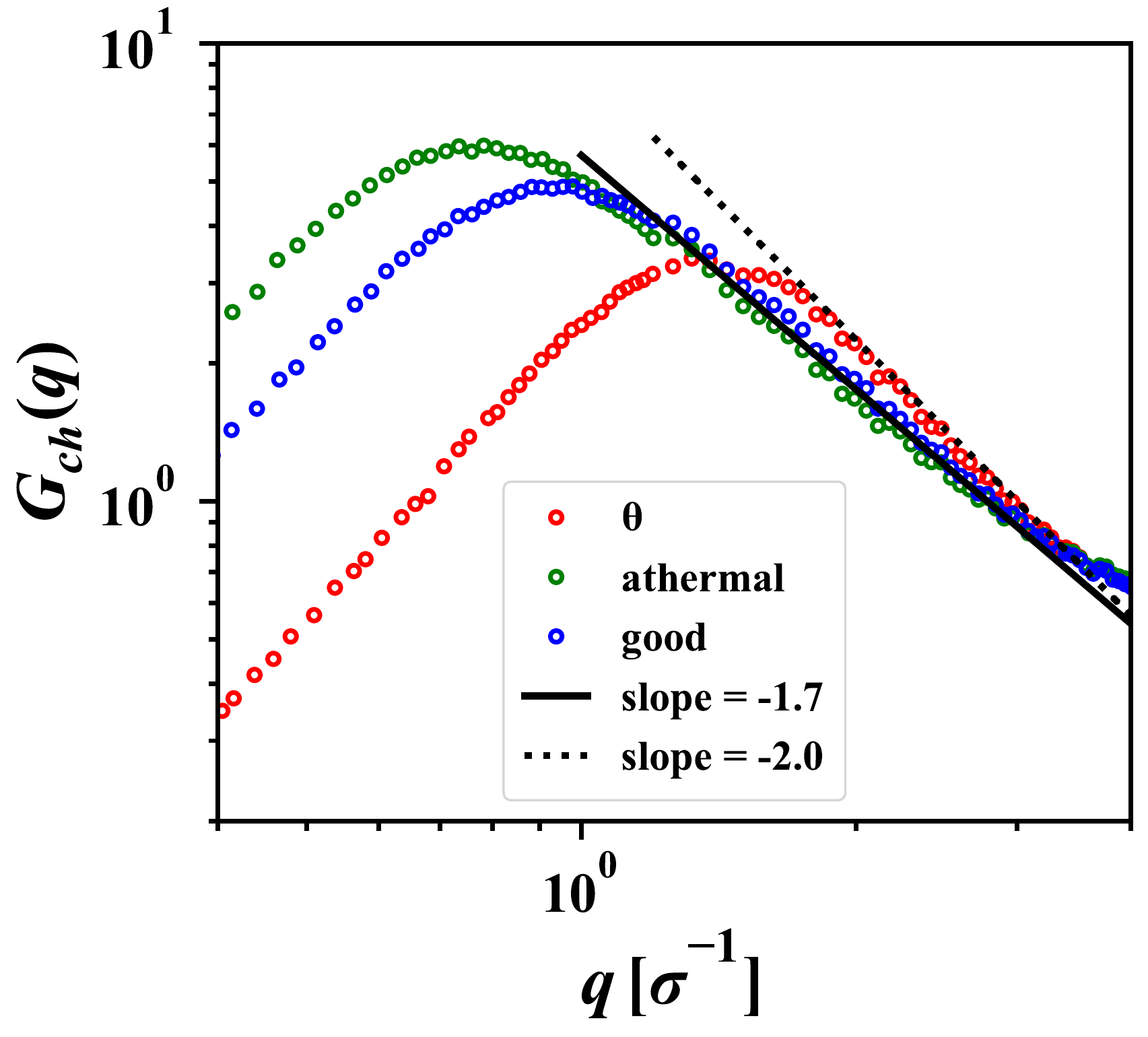}
\caption{Charge structure factor of salt-free PECs at different solvent quality from simulations. In $\theta$ solvent, the charge structure factor decay with $G_{ch} \sim q^{-2}$ at high $q$ regime, as shown by the dotted black line, indicating the ideal chain conformation in charged blobs. In athermal and good solvent, the charge structure factor decay with $G_{ch} \sim q^{-1.7}$ at high $q$ regime, as shown by the solid black line, indicating the swollen chain conformation in charged blobs.}
\label{fig:Gch_slope}
\end{figure}

\begin{figure}
    \centering
        \includegraphics[width=0.475\textwidth]{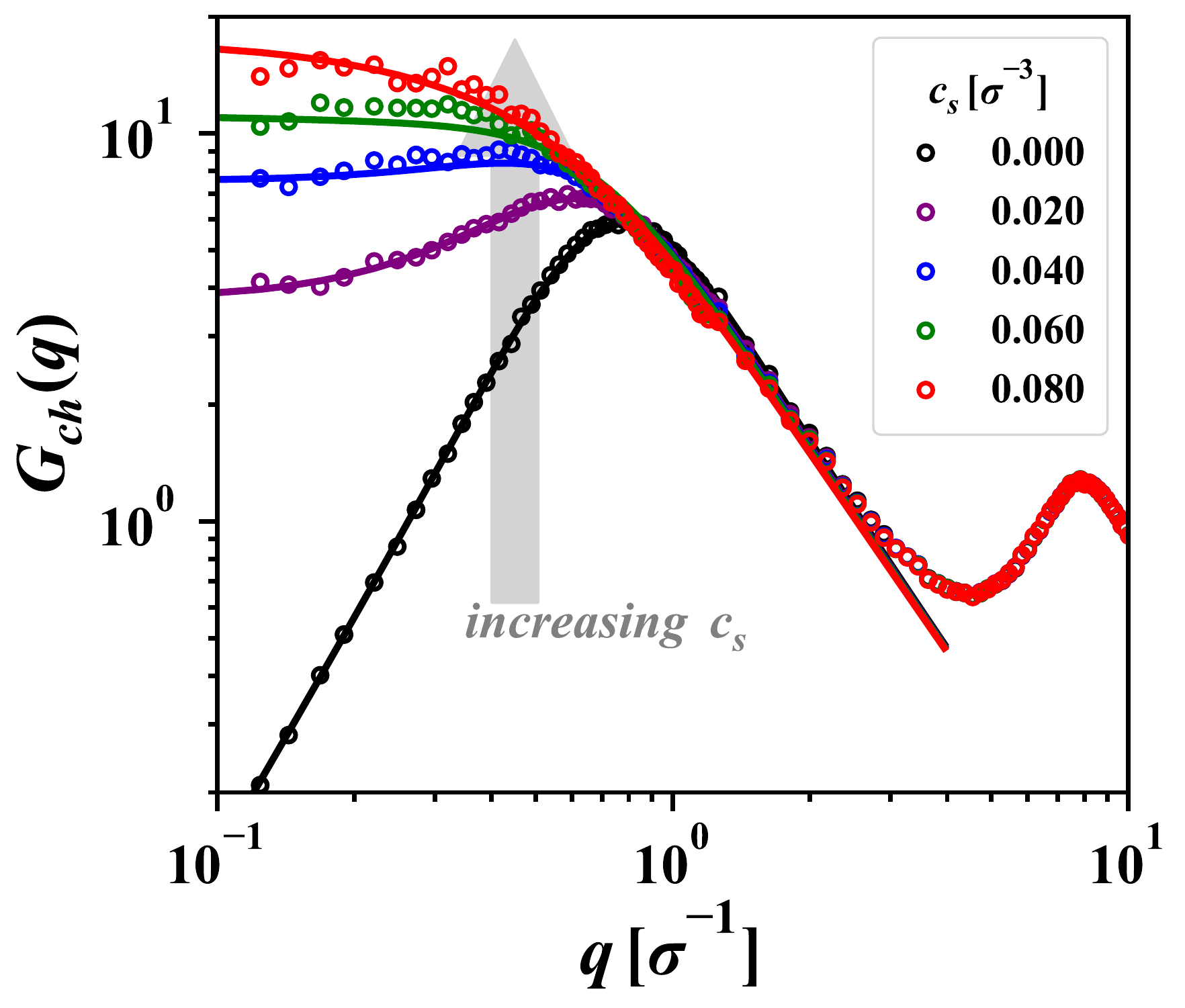}
        \includegraphics[width=0.475\textwidth]{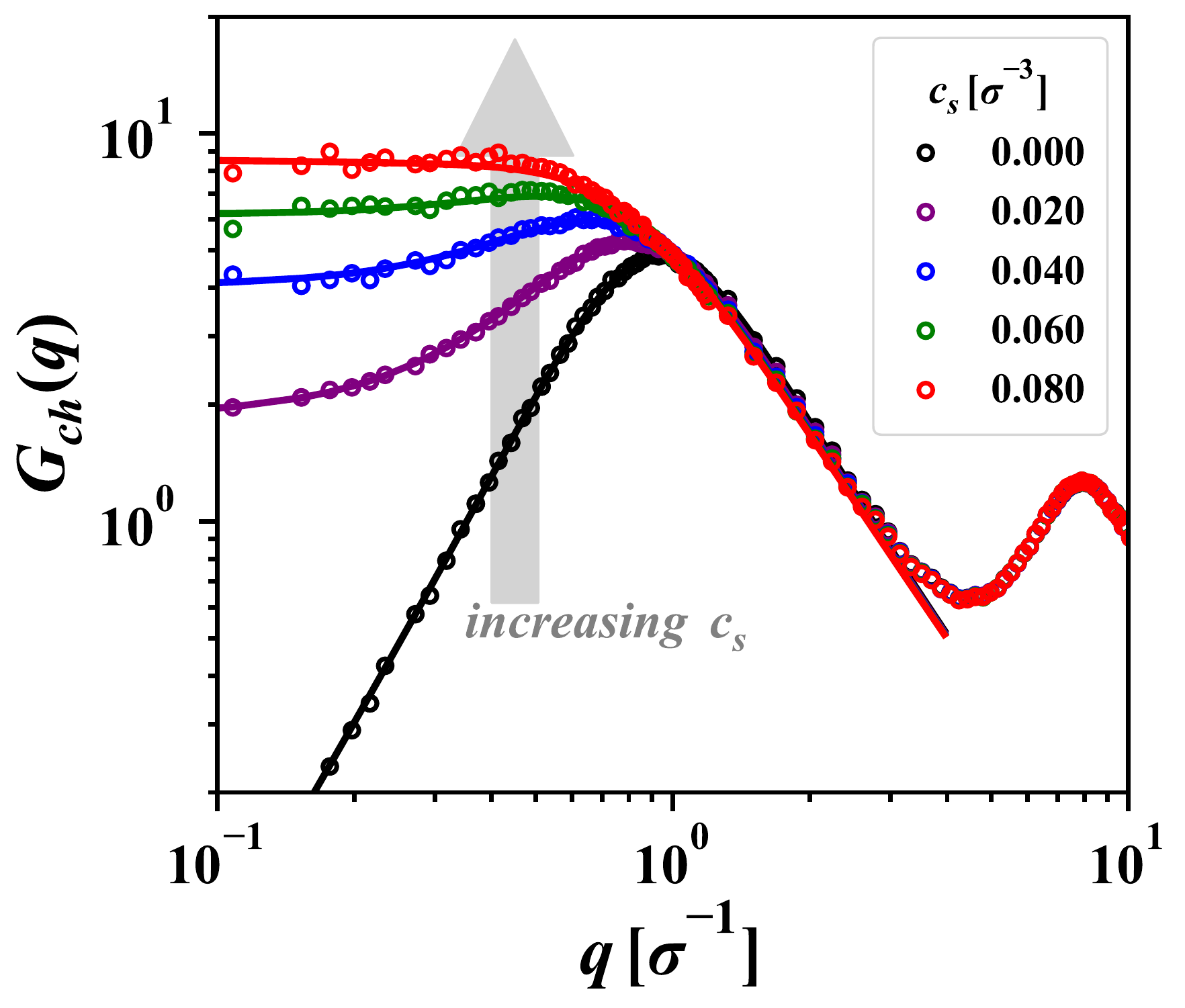}
        \includegraphics[width=0.475\textwidth]{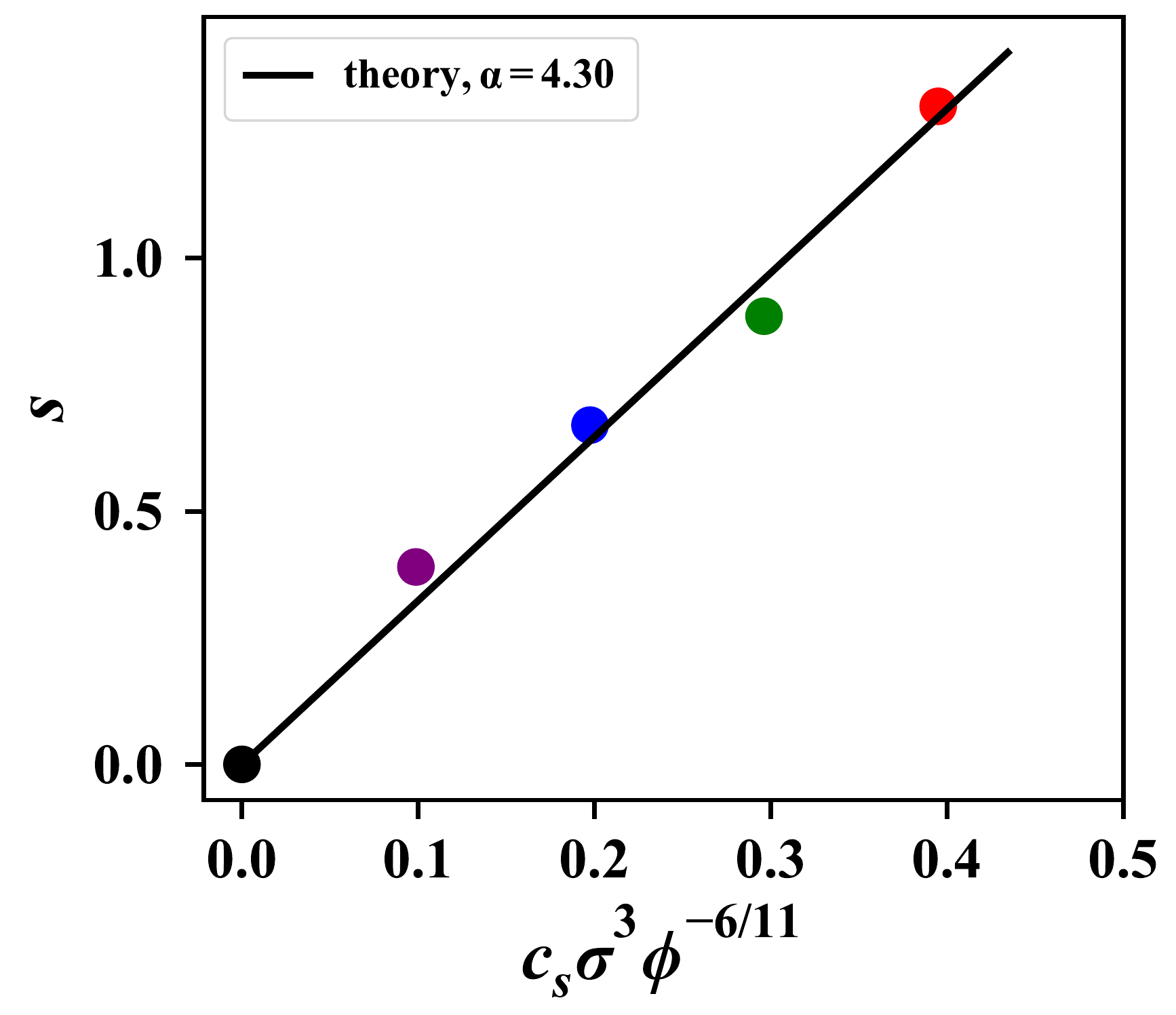}
        \includegraphics[width=0.475\textwidth]{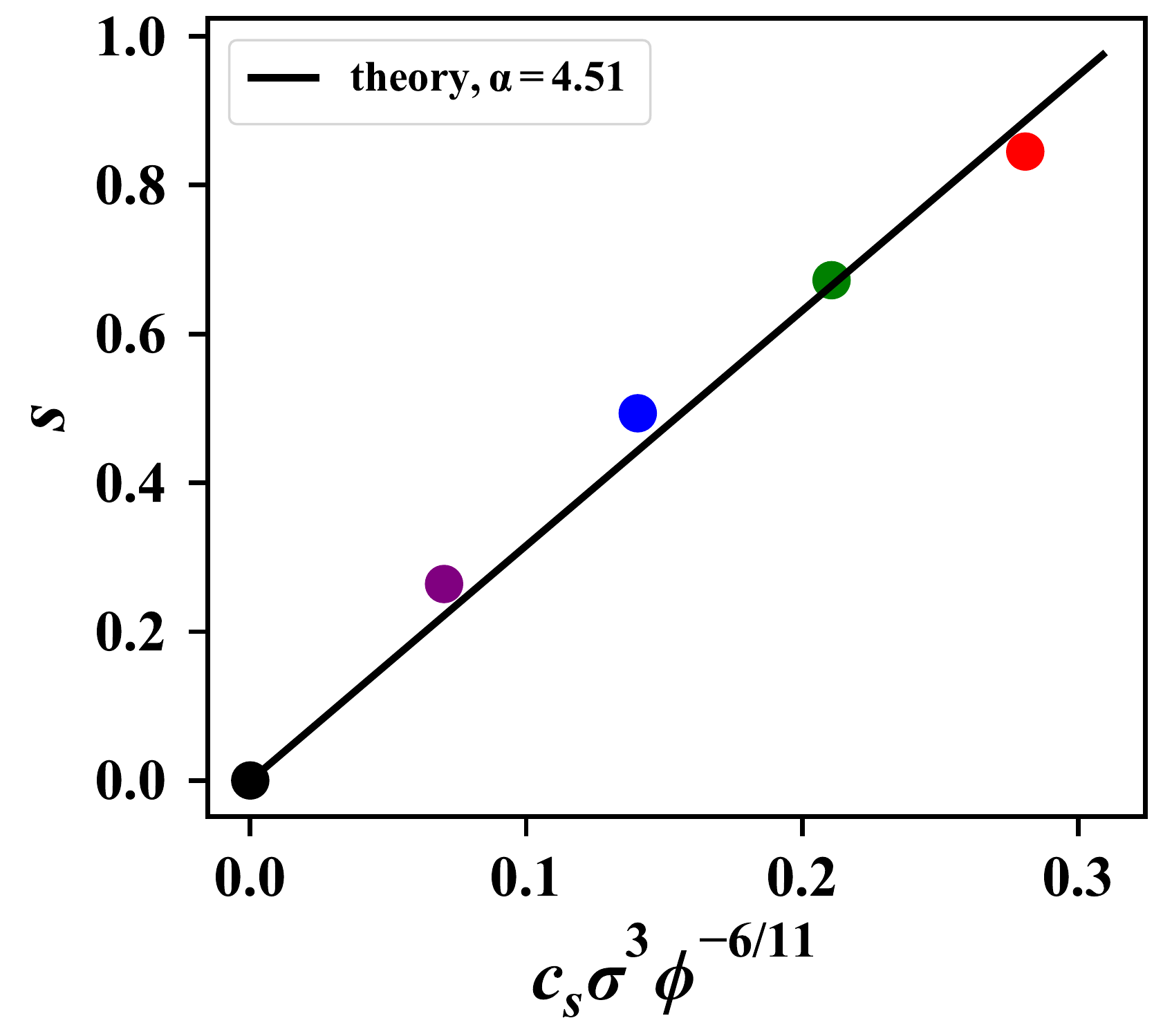}
    \caption{(Top) Charge structure factor of PECs in an athermal (top left) and a good (top right) solvent under the constant polymer concentration with varying salt concentrations. Solid lines are the best theoretical fits, with the functional form given by: $G_{ch} (Q) \propto \left( 1/(Q^2 + s) + Q^{5/3} + R \right)^{-1}$. (Bottom) Dependence of the parameter $s$ on salt concentration $c_s$ in athermal (bottom left) and good (bottom right) solvents is linear, in agreement with the RPA predictions, eq.~\ref{s_a}. Solid lines are the best theoretical fits of eq.~\ref{s_a} in the Supporting Information, with $\alpha$ being the only fitting parameter.} 
    \label{fig:athermal_good}
\end{figure}

\newpage
\section{Experimental Details}

\subsection{Materials and Methods}
All reagents were obtained from Sigma Aldrich and used as received unless otherwise specified. Ethylene oxide (>99\%, lecture bottle) was obtained from Praxair, Inc. Poly(AGE$_{55}$-\emph{stat}-EO$_{128}$) was synthesized in \cite{neitzel-2021}. For the current work this identical neutral precursor was used to generate more copolyelectrolyte material. The deuterated copolyelectrolytes were synthesized as reported previously \cite{neitzel-2021} except that tetrahydrofuran (HPLC grade) from the solvent purification system (MBraun SPS-800) was further purified via vacuum transfer from sodium/benzophenone prior to polymerization. In the current work, all thiol-ene click reactions were conducted in 4/1 dimethylformamide/water as the solvent. Proton nuclear magnetic resonance ($^1$H NMR) spectroscopy was performed using a Bruker Avance III 400 MHz spectrometer collecting 64 scans at a relaxation delay (d1) of 10s. Signals were referenced to the internal standards tetramethylsilane (TMS) in CDCl$_3$ and 3-(trimethylsilyl) propionic-2,2,3,3-$d_4$ acid, sodium salt in D$_2$O. $^1$H NMR spectroscopic data was processed with MestrReNova. Neutral copolymers were characterized by size exclusion chromatography (SEC). Poly(AGE$_{55}$-\emph{stat}-EO$_{128}$) was analyzed with 0.01 M sodium bromide in dimethylformamide as the eluent (flow rate = 0.3 mL/min, T = 50 ℃) using a Tosoh EcoSEC instrument equipped with three chromatography columns (Tosoh SuperAW3000, SuperAW4000, and guard column) and a refractive index detector. Poly(AGE$_{64}$-\emph{stat}-$d_4$-EO$_{142}$) was analyzed in tetrahydrofuran as the mobile phase (flow rate = 1.0 mL/min, T = 25 °C) using a Shimadzu Prominance LC system equipped with two Agilent PLgel 5 micron Mixed-D columns + a guard column, Wyatt DAWN HELEOS II MALS (658 nm laser) and Wyatt Optilab T-rEX refractive index (RI) detectors.

\subsection{Contrast matching}

Contrast matching conditions were calculated (https://www.ncnr.nist.gov/resources/activation/) and measured experimentally for polyelectrolyte solutions (50 mg/mL) using 4 H$_2$O/D$_2$O ratios. Results are displayed in Table S1 and Figure S7.

\begin{table} [h]
\scriptsize
\centering
\caption{Experimentally Measured and theoretically calculated copolyelectrolyte contrast matching conditions and scattering length densities}
\label{table:exp2}%
\begin{tabular}{|c|c|c|c|c|}
\hline
\multirow{2}{*}{Chemical} & 
Experimentally measured & 
Calculated  & 
Experimentally measured  & 
Calculated 
\\ 
&
vol\% \ch{H2O} / vol\% \ch{D2O} &
vol\% \ch{H2O} / vol\% \ch{D2O} &
SLD [ $10^{-6} \angstrom^{-2}$ ] &
SLD [$10^{-6} \angstrom^{-2} $]
\\ \hline
H$_2$O &
\multicolumn{2}{|c|}{1} &
\multicolumn{2}{|c|}{-0.56}
\\  \hline
D$_2$O & 
\multicolumn{2}{|c|}{0} &
\multicolumn{2}{|c|}{6.40} 
\\  \hline
 poly(Sulf$_{55}$-\emph{stat}-EO$_{128}$) & 71.9/28.1 & 81.3/18.7 & 1.396  & 0.741
\\  \hline
 poly(Am$_{ox}$$_{55}$-\emph{stat}-EO$_{128}$) & 76.8/23.2 & 83.3/16.7   & 1.055 & 0.599
\\  \hline
 poly(Sulf$_{64}$-\emph{stat}-$d_4$-EO$_{142}$) & 57.9/42.1 & 55.8/44.2  & 2.370  & 2.518
\\  \hline
 poly(Am$_{ox}$$_{64}$-\emph{stat}-$d_4$-EO$_{142}$) & 48.6/51.4 & 53.6/46.4  & 3.017 & 2.671
\\  \hline
\end{tabular}%
\end{table}

\newpage
 
\section{Copolymers and Copolyelectrolytes Characterization Data}


\begin{figure} [H]
\centering
\includegraphics[width=1\linewidth]{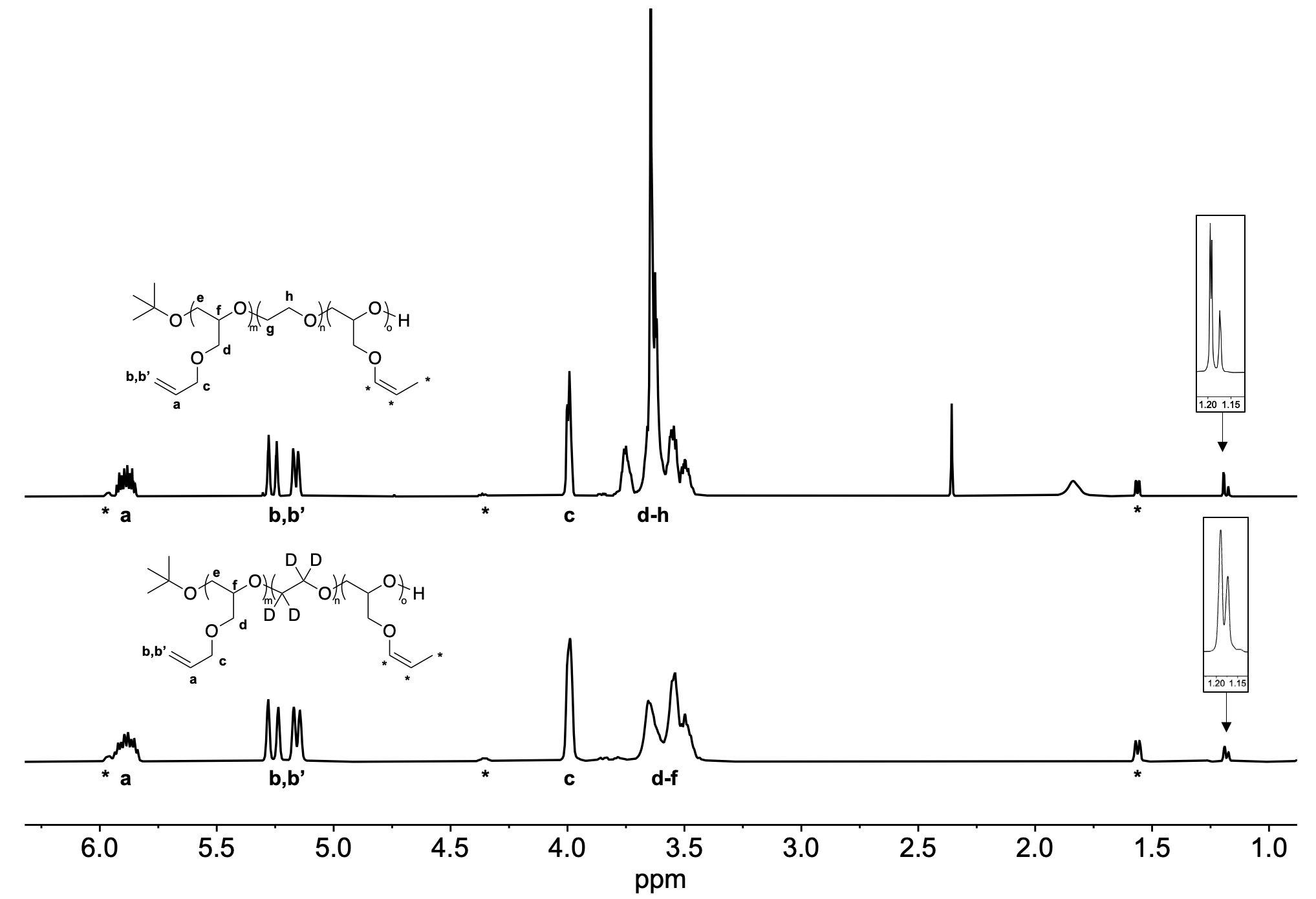}
\caption{Overlaid $^1$H NMR spectra (CDCl$_3$) of Poly(AGE$_{55}$-\emph{stat}-EO$_{128}$) and Poly(AGE$_{64}$-\emph{stat}-$d_4$-EO$_{142}$) copolymers synthesized to produce the homologous copolyelectrolytes studied in this work. The end group signal derived from tert-butoxide is magnified and corresponds to 9 protons. Minor signals (*) are attributed to AGE olefins that have isomerized to the internal cis-alkene. The fraction of isomerized alkene units are: Poly(AGE$_{55}$-\emph{stat}-EO$_{128}$)$\rightarrow$ 6.4\% and Poly(AGE$_{64}$-\emph{stat}-$d_4$-EO$_{142}$)$\rightarrow$8.7\%.
 }
\label{fig:blobs}
\end{figure}


\begin{figure}
\centering
\includegraphics[width=1\linewidth]{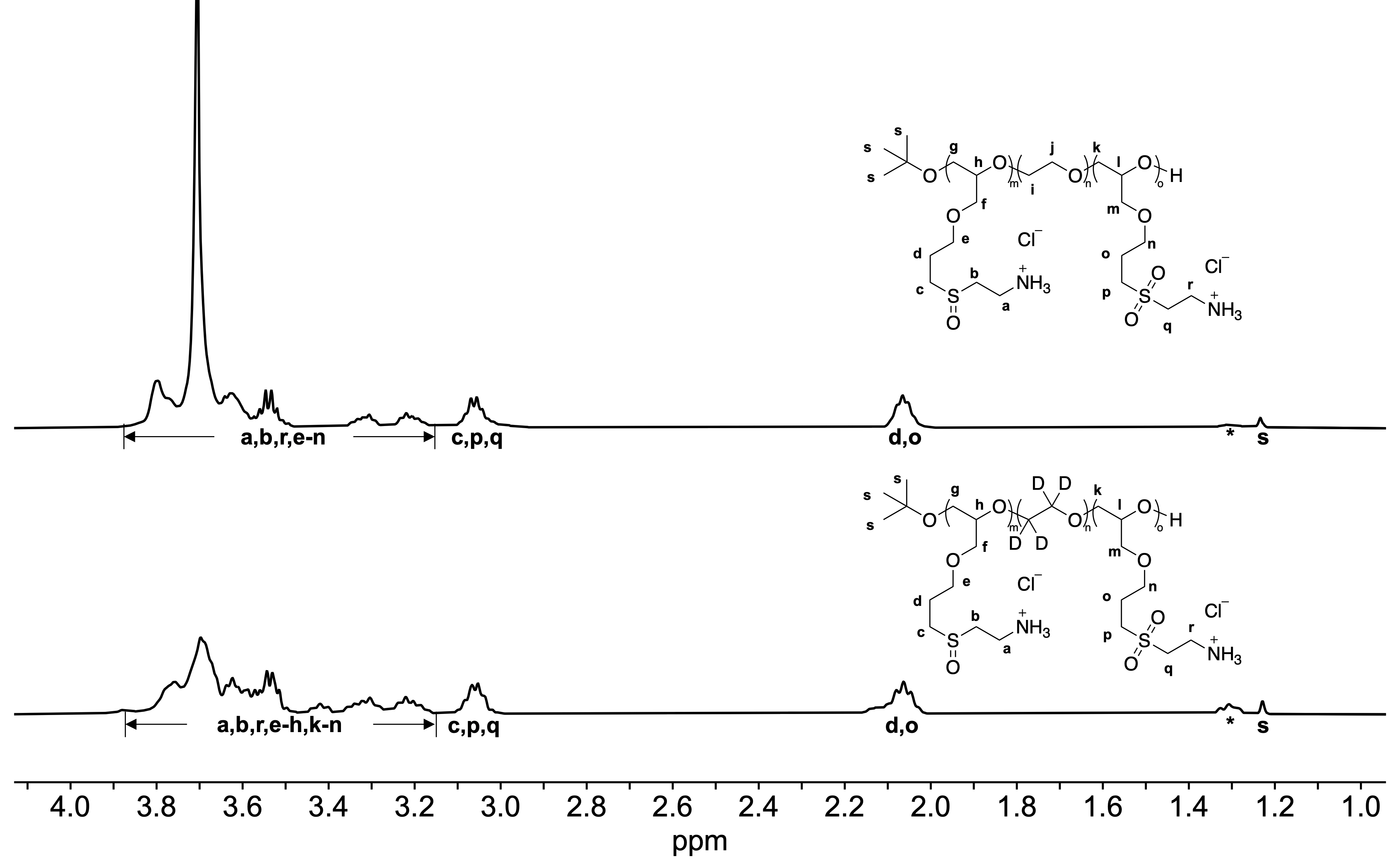}
\caption{Overlaid $^1$H NMR spectra (D$_2$O) of poly(Am$_{ox 55}$-\emph{stat}-EO$_{128}$) and  poly(Am$_{ox 64}$-\emph{stat}-$d_4$-EO$_{142}$) copolycations synthesized from the neutral copolymers in Figure S2 via thiol-ene click functionalization with cysteamine hydrochloride and subsequent oxidation with 2 eq. H$_2$O$_2$ relative to moles of sulfur. Minor signals (*) are attributed to products of thiol-ene clicks with isomerized AGE olefins.}
\label{fig:blobs}
\end{figure}


\begin{figure}
\centering
\includegraphics[width=1.0\linewidth]{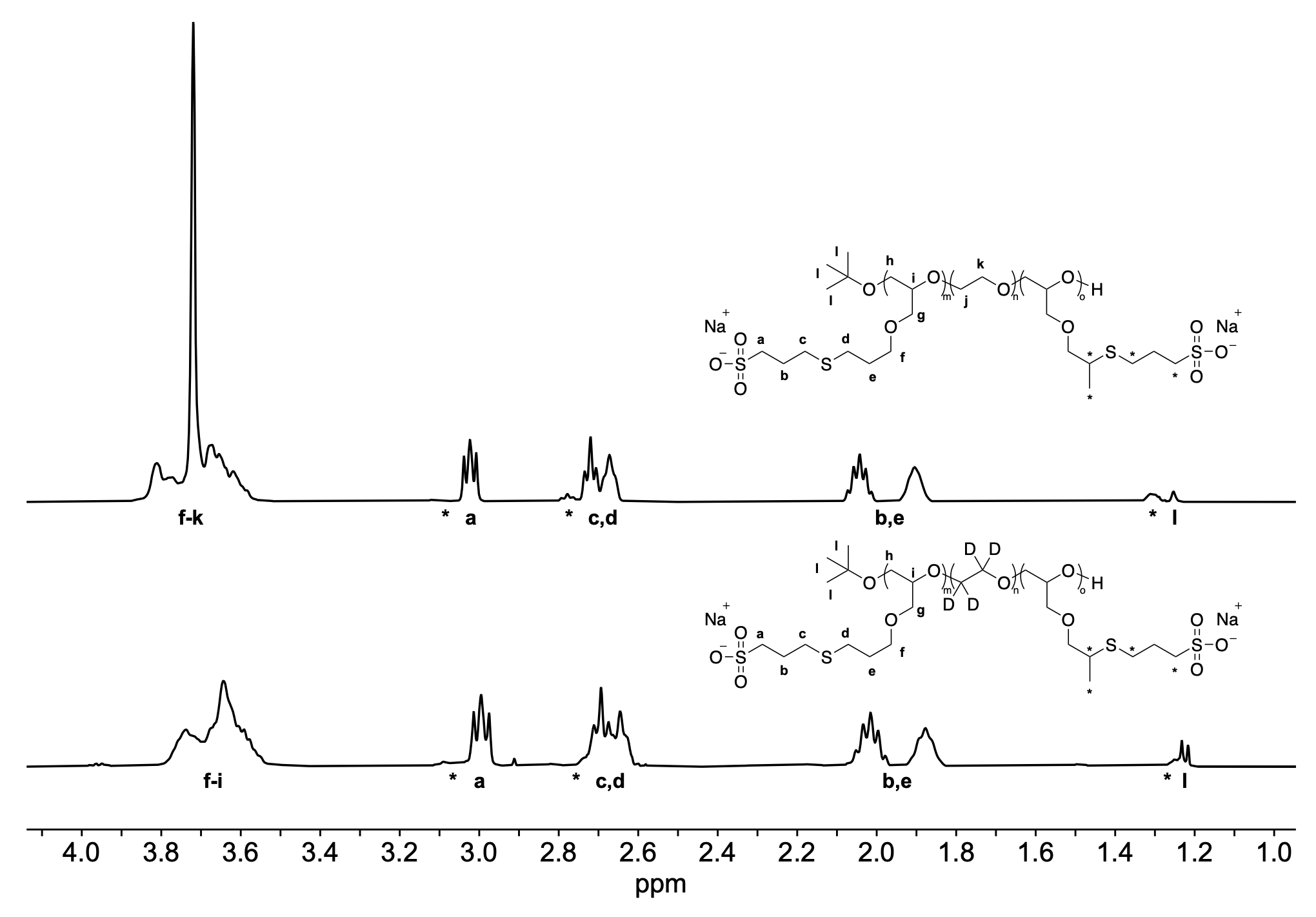}
\caption{Overlaid $^1$H NMR spectra (D$_2$O) of poly(Sulf$_{55}$-\emph{stat}-EO$_{128}$) and  poly(Sulf$_{64}$-\emph{stat}-$d_4$-EO$_{142}$) copolyanions synthesized from the neutral copolymers in Figure S1 via thiol-ene click functionalization with sodium 3-mercapto-1-propanesulfonate. Minor sigals (*) are attributed to products of thiol-ene clicks with isomerized AGE olefins.
 }
\label{fig:blobs}
\end{figure}

\begin{figure}
\centering
\includegraphics[width=0.8\linewidth]{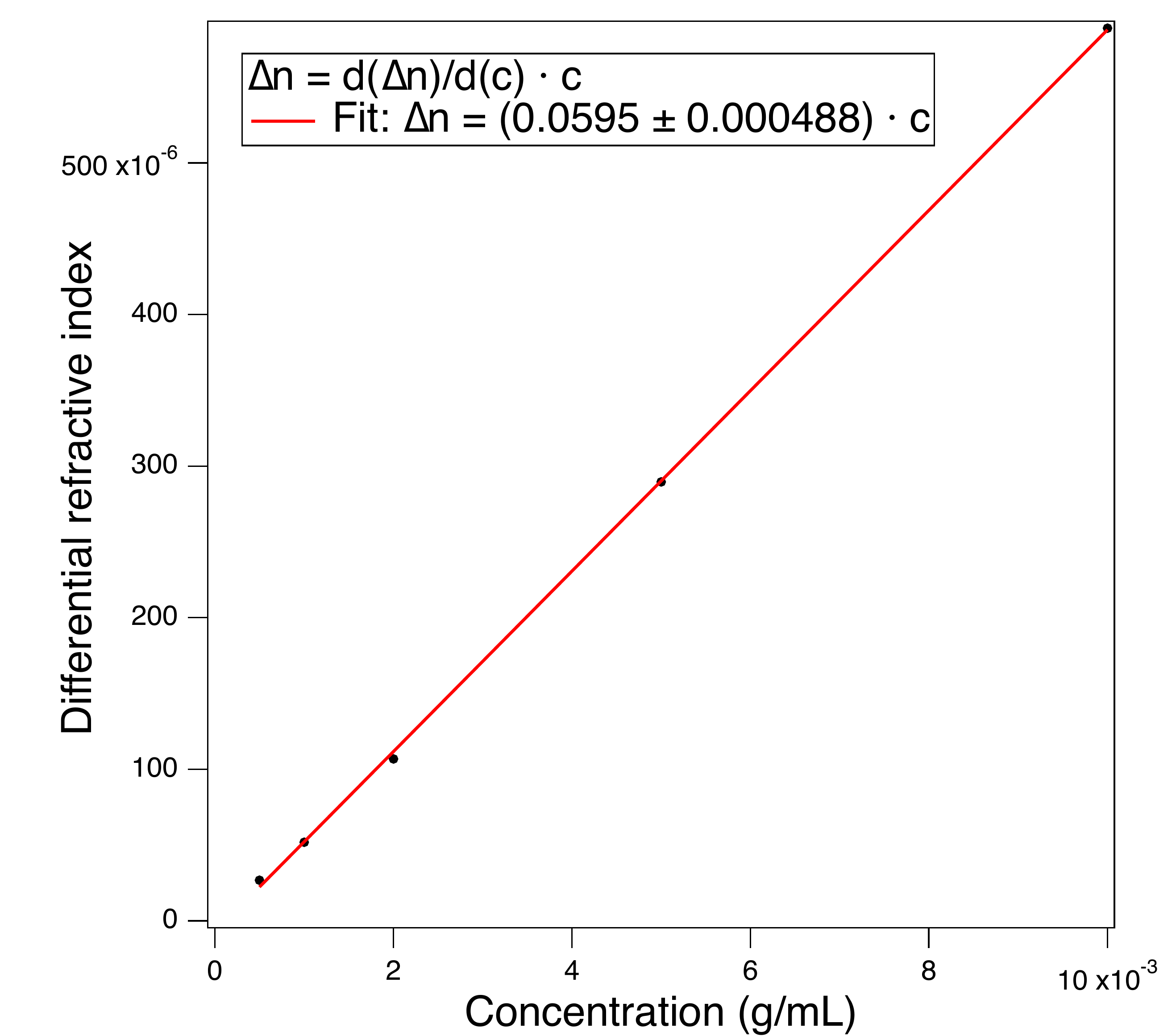}
\caption{Batch injections with increasing concentrations of poly(AGE$_{64}$-\emph{stat}-$d_4$-EO$_{142}$) in THF solution were performed to determine the change in differential refractive index ($\Delta n$) as a function of concentration ($c$). The refractive index increment d($\Delta n$)/dc is calculated from the slope of the graph of differential refractive index ($\Delta n$) vs concentration ($c$).
}
\label{fig:blobs}
\end{figure}


\begin{figure}
\centering
\includegraphics[width=0.8\linewidth]{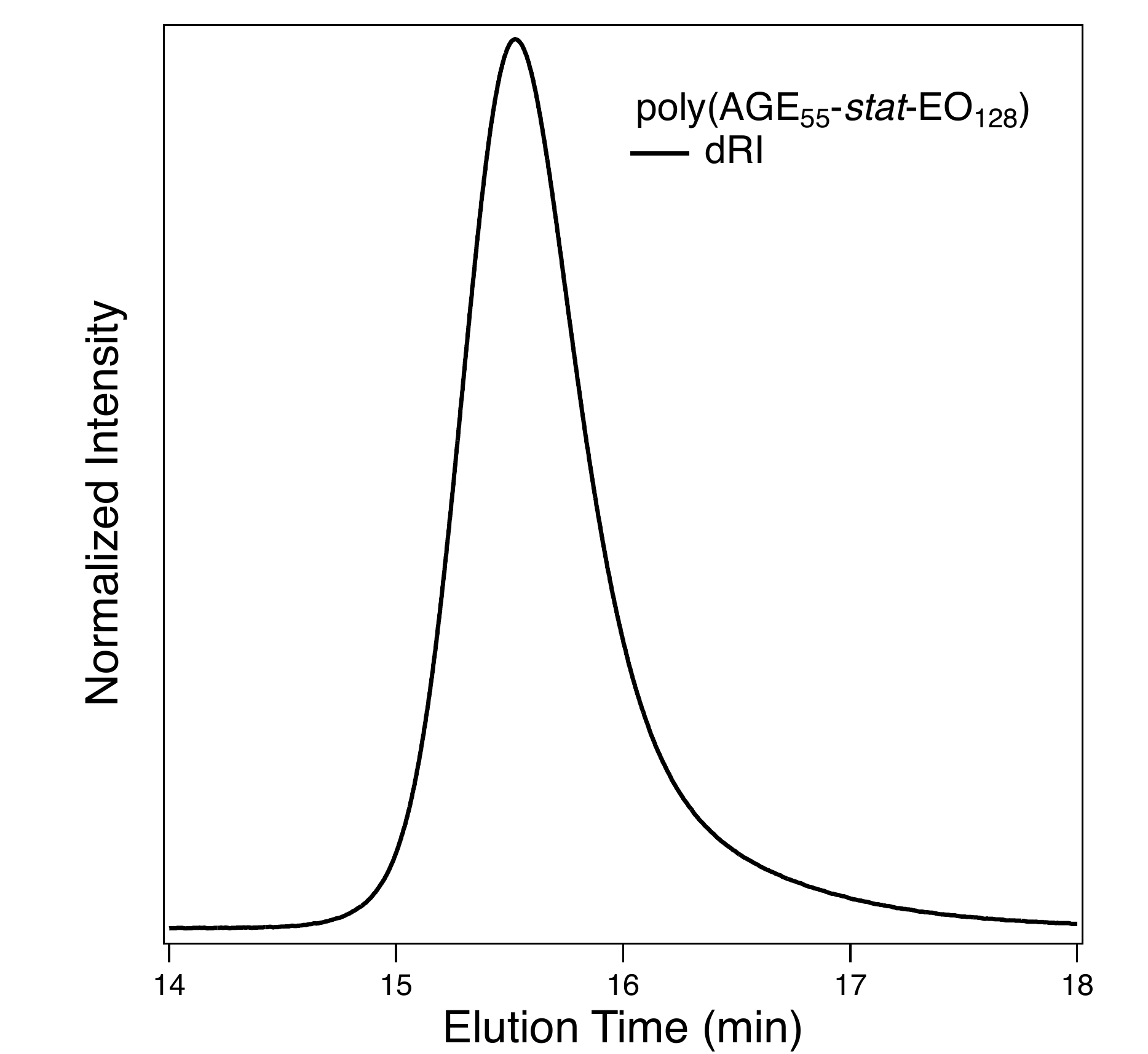}
\caption{DMF SEC trace of Poly(AGE$_{55}$-\emph{stat}-EO$_{128}$). The sample was prepared in 0.01 M sodium bromide in DMF and eluted at a flow rate of 0.3 mL/min. Dispersity was calculated from the differential refractive index signal which gave  {\it Đ} = 1.15.} 
\label{fig:blobs}
\end{figure}


\begin{figure}
\centering
\includegraphics[width=0.8\linewidth]{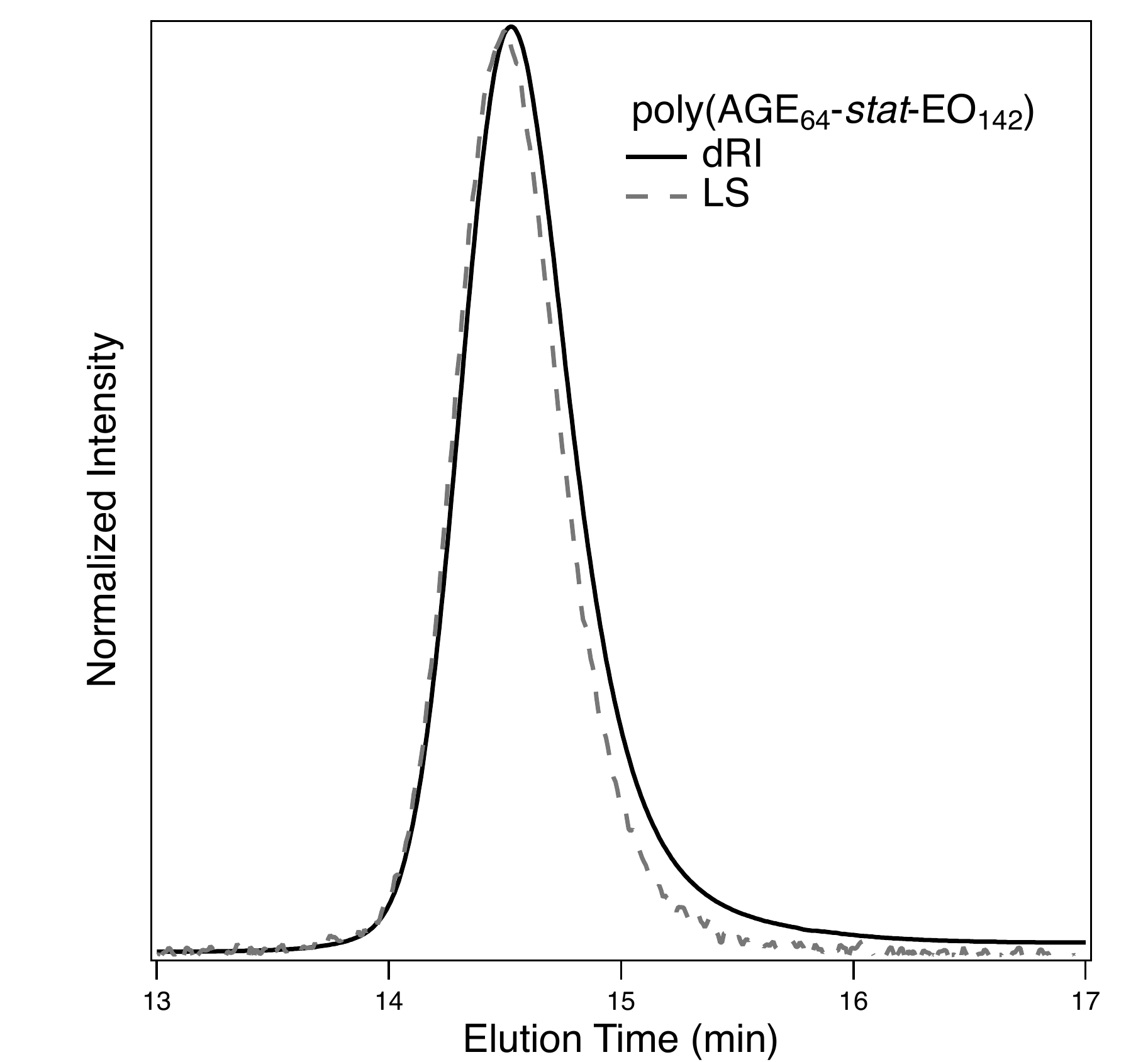}
\caption{THF SEC trace of Poly(AGE$_{64}$-\emph{stat}-$d_4$-EO$_{142}$). The sample was prepared at a concentration of 5mg/mL in THF and eluted at a flow rate of 1.0 mL/min. Molar mass analysis by MALLS provided {\it M$_w$} = 13.77 kg/mol. Dispersity calculated from the light scattering signal gave  {\it Đ} = 1.05. Dispersity calculated from the differential refractive index signal gave  {\it Đ} = 1.07.} 
\label{fig:blobs}
\end{figure}

\begin{figure}
\centering
\includegraphics[width=0.4\linewidth]{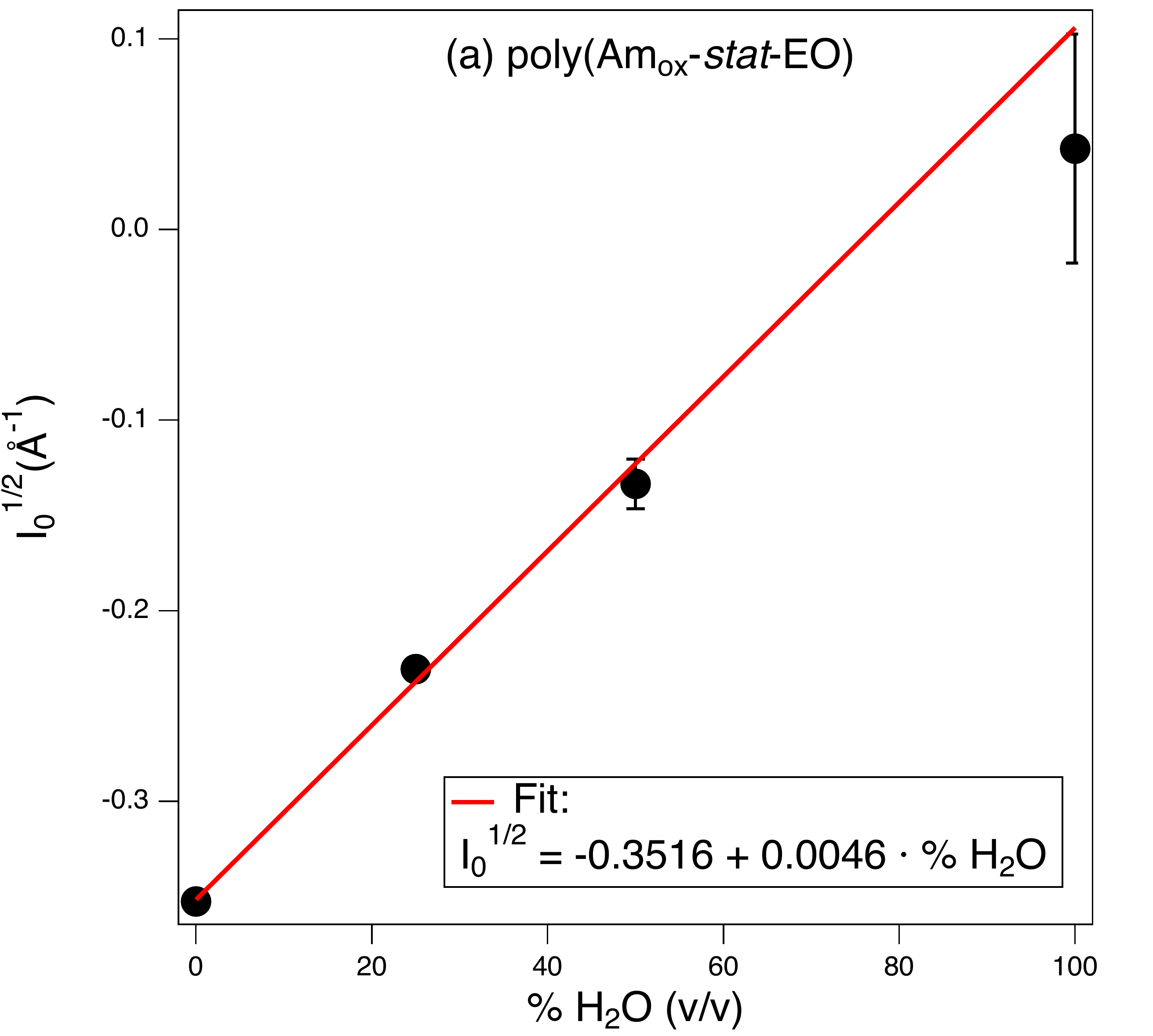}
\includegraphics[width=0.4\linewidth]{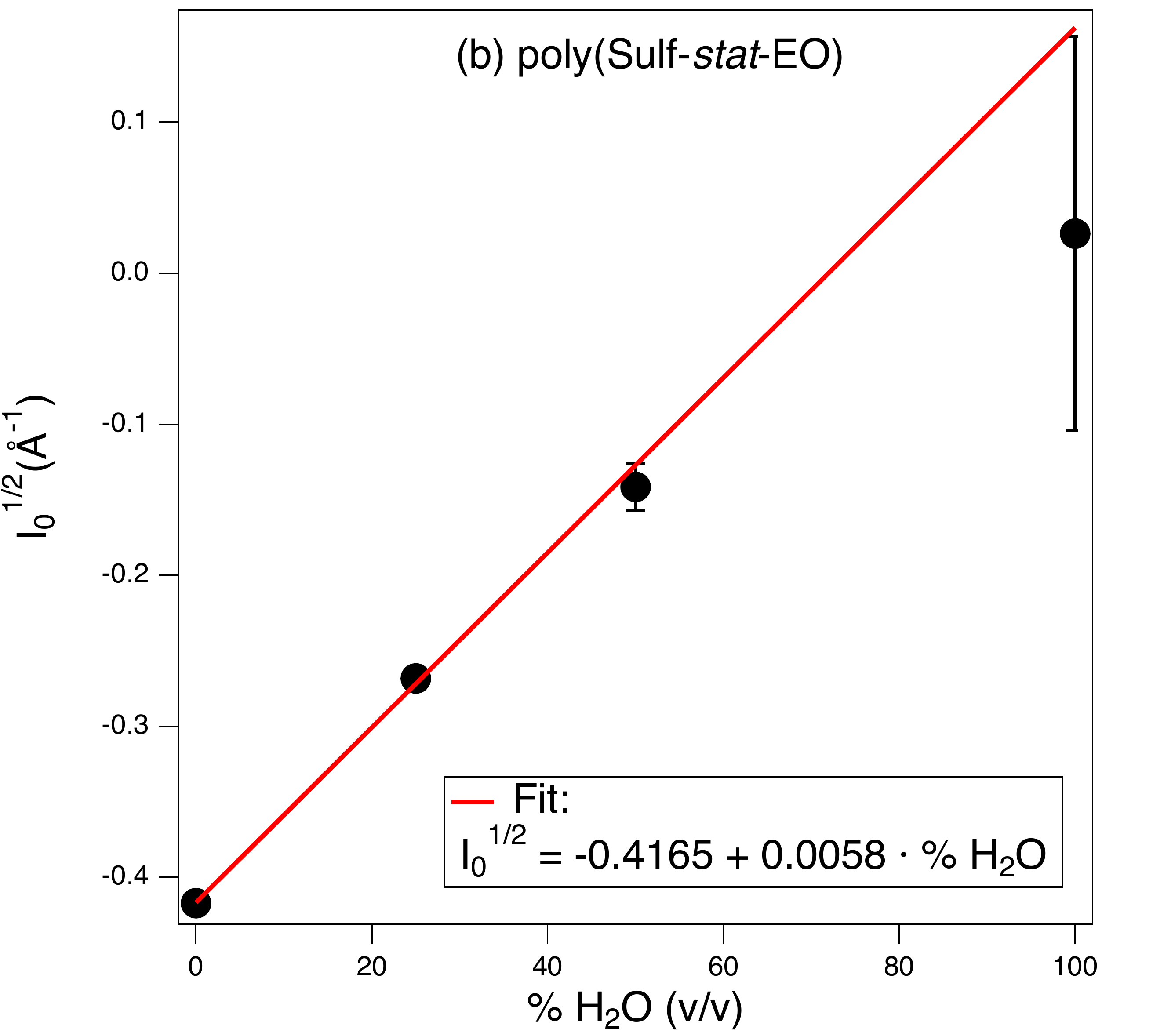}
\includegraphics[width=0.4\linewidth]{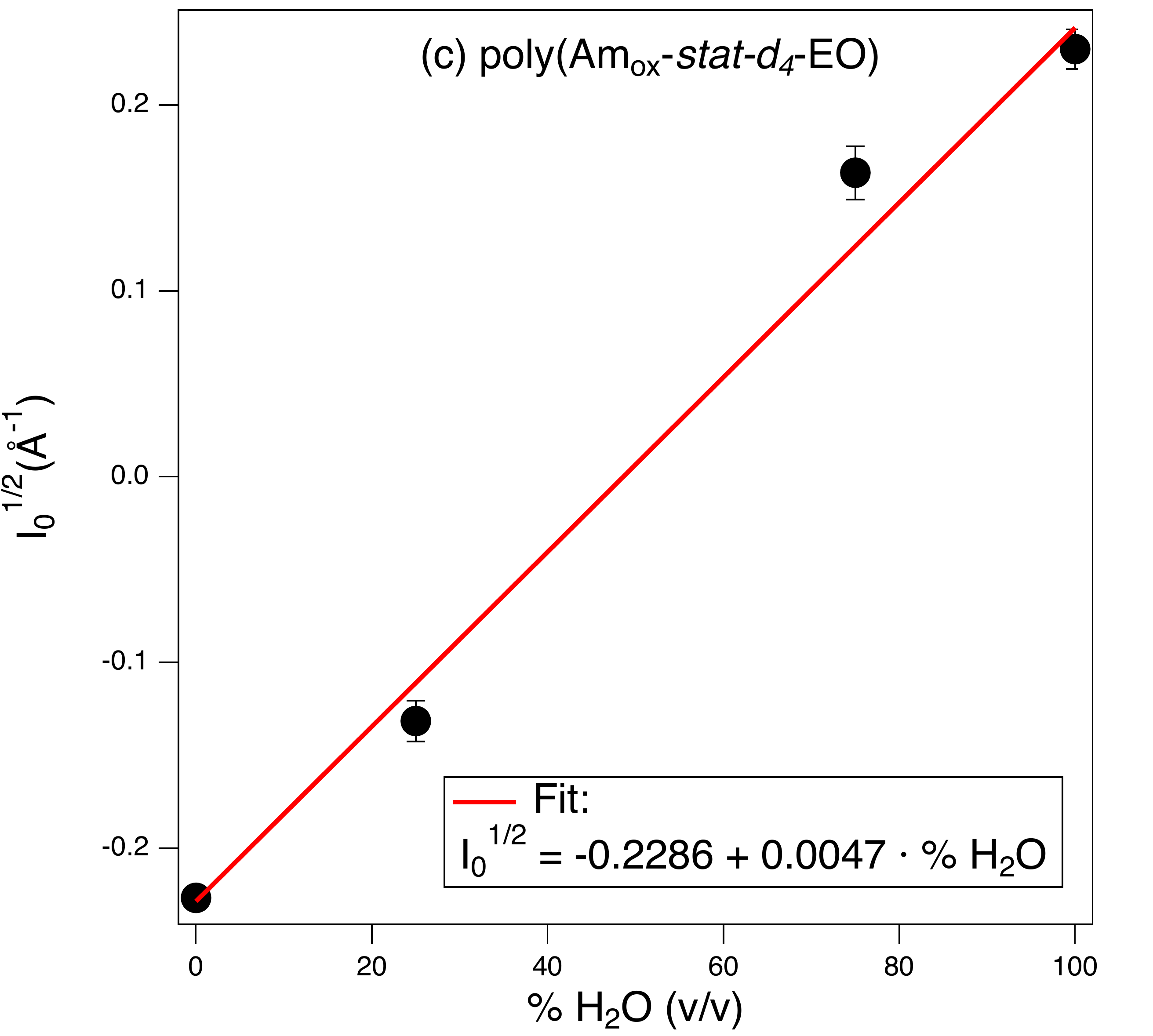}
\includegraphics[width=0.4\linewidth]{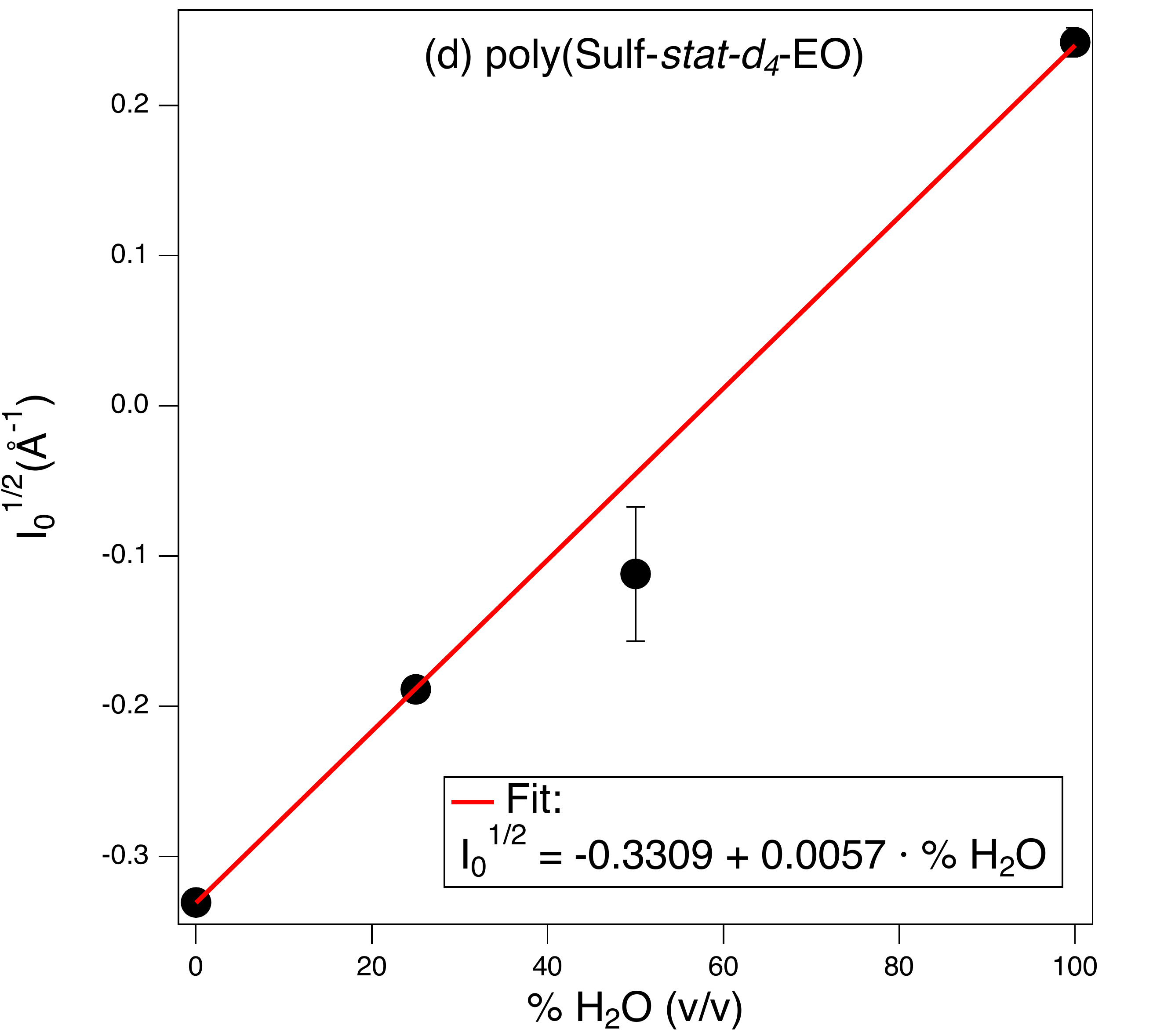}
\caption{Contrast matching conditions were determined experimentally by measuring
the neutron scattering intensity ($I_0^{1/2}$) for polyelectrolyte solutions (50 mg/mL) with: $\phi_{\ch{H2O}}$ = 0, 0.25, 0.75, and 1. The \% \ch{H2O} were calculated at  $I_0^{1/2}$= 0 to give:
(a) poly(Am$_{ox}$-\emph{stat}-EO), \% H$_2$O=76.8; 
(b) poly(Sulf-\emph{stat}-EO), \% H$_2$O=71.9; 
(c) poly(Am$_{ox}$-\emph{stat}-$d_4$-EO), \% H$_2$O=48.6; 
(d) poly(Sulf-\emph{stat}-$d_4$-EO), \% H$_2$O=57.9. 
}
\label{fig:blobs}
\end{figure}